\documentclass[aps,prd,groupedaddress,showpacs,superscriptaddress]{revtex4-1}

\usepackage{amsmath}
\usepackage{graphicx}
\usepackage{amssymb}
\usepackage{bm}
\usepackage{array}
\usepackage{hyperref}
\usepackage{slashed}
\usepackage{braket}
\usepackage{gensymb}

\begin{document}


\title{Present status on experimental search for pentaquarks}

\newcommand*{\PKU}{School of Physics and State Key Laboratory of Nuclear Physics and
Technology, Peking University, Beijing 100871,
China}\affiliation{\PKU}
\newcommand*{\CIC}{Collaborative Innovation Center of Quantum Matter, Beijing, China}\affiliation{\CIC}
\newcommand*{\CHEP}{Center for High Energy
Physics, Peking University, Beijing 100871,
China}\affiliation{\CHEP}

\author{Tianbo Liu}\affiliation{\PKU}
\author{Yajun Mao}\affiliation{\PKU}
\author{Bo-Qiang Ma}\email{mabq@pku.edu.cn}\affiliation{\PKU}\affiliation{\CIC}\affiliation{\CHEP}

\date{\today}

\begin{abstract}
It has been ten years since the first report for a positive strangeness pentaquark-like baryon state. However the existence of the pentaquark state is still controversial. Some contradictions between the experiments are unsolved. In this paper we review the experimental search for the pentaquark candidates $\Theta^+$, $\Theta^{++}$, $\Xi^{--}$, $\Theta_c^0$ and $N^*$ in details. We review the experiments with positive results and compare the experiments with similar conditions but opposite results.
\end{abstract}



\maketitle

\section{Introduction}

Hadrons are the bound states of the strong interaction which is described by the quantum chromodynamics (QCD) in the framework of Yang-Mills gauge theory. One of the main goals of the hadron physics is to understand the composition of hadrons in terms of quarks and gluons. The quark model is proved successful in classifying the mesons and baryons as $q\bar{q}$ and $qqq$ composite systems. Almost all the well established mesons can be described as a quark-antiquark state except some mesons with exotic quantum numbers which are impossible for a $q\bar{q}$ system, but no experimental evidence is reported for exotic baryons which are inconsistent with the $qqq$ configuration until the beginning of this century. Theoretically, the QCD does not forbid the existence of the hadrons with the other configurations, such as the glueballs, the hybrids and the multiquarks. In this review, we focus on the pentaquarks. If the pentaquark really exists, it will provide a new stage to test the QCD in the nonperturbative region and to investigate the properties of the strong interaction.

In the quark model language, the pentaquark is a hadron state with four valence quarks and one valence antiquark as $|qqqq\bar{q}\rangle$~\cite{GellMann:1964nj,Jaffe:1976ih,Aerts:1977rw,deCrombrugghe:1978hi}. Because the pentaquark can decay to a three-quark baryon and a quark-antiquark meson, its width was suggested to be wide~\cite{Jaffe:1976ig,Strottman:1979qu}, but it was predicted to have a narrow width due to its particular quark structure~\cite{Lipkin:1998pb,Diakonov:1997mm}. In principle, any baryon may have the five-quark contents, and experiments have revealed the important role of the intrinsic sea quarks in understanding the structure of the proton. On the other hand, the pentaquark state may also mix with the corresponding three-quark state or hybrid state, so the situation is much more complicated. The pentaquark is easier to be identified if it has no admixture with any three-quark state,
i.e., if the
flavor of the anti-quark $\bar{q}$ in the $|qqqq\bar{q}\rangle$ state is different from any of the other four
quarks~\cite{Gao:1999ar}. Early experiments in 1960's and 1970's were performed to search for a baryon with positive strangeness as the pentaquark candidate referred to as the $Z^*$~\cite{Anderson:1969ng}, but no enhancements were found.

This field developed rapidly on both the experimental and the theoretical aspects in the last decade since the first report for a positive strangeness pentaquark-like baryon, referred to as the $\Theta^+$, by the LEPS Collaboration~\cite{Nakano:2003qx}. Its mass and width are closed to the prediction of the chiral soliton model~\cite{Diakonov:1997mm}. This particle was quickly reported in subsequent experiments by some groups, and many theoretical models were applied to understanding this particle and to predicting other pentaquarks, such as the diquark cluster model~\cite{Jaffe:2003sg,Jaffe:2003ci,Narodetskii:2003ft}, the diquark-triquark model~\cite{Karliner:2003dt}, the constituent quark model~\cite{Stancu:2003if,Capstick:2003iq,Carlson:2003pn,Cheung:2003de}, the chiral quark model~\cite{Glozman:2003sy,Huang:2003we}, the bag model~\cite{Hosaka:2003jv}, the meson-baryon binding~\cite{Itzhaki:2003nr,Kahana:2003rv,LlanesEstrada:2003us}, the QCD sum rules~\cite{Zhu:2003ba,Sugiyama:2003zk}, the lattice QCD~\cite{Csikor:2003ng,Sasaki:2003gi,Mathur:2004jr,Chiu:2004gg,Ishii:2004qe,Takahashi:2004sc} and the chiral soliton model in new versions~\cite{Diakonov:2003jj,Wu:2003nw}. Unfortunately, many experiments found no signals for this particle. What is worse, the signals observed in the early experiments by some groups disappeared when including the new data with high statistics. However, some groups reconfirmed their observations for this particle with very high statistical significance in their updated experiments. So even the existence of the pentaquark is a mystery. The production mechanism and the analysis method should be investigated in details.

Recently, a charged charmonium-like meson $Z_c(3900)$ was observed by BES~\cite{Ablikim:2013mio} and Belle~\cite{Liu:2013dau}. It is a suggestive evidence for the existence of the multiquark meson. This arouses much interest on the study of the multiquark states. In this paper, we review the experimental search for the pentaquark states. In Sect. II and III, we concentrate on the searches for the $\Theta^+$ with positive or negative results. In Sect. IV, we focus on a non-strange pentaquark candidate. In Sect. V, the other observed pentaquark candidates are presented. Then we discuss the results in Sect. VI and a brief summary is contained in the last section.

\section{Experiments with positive results for $\Theta^+$}

The pentaquark candidate $\Theta^+$ was widely discussed and searched for since the first report on the experimental observation by the LEPS Collaboration~\cite{Nakano:2003qx}. The Skyrme's idea that baryons are solitons~\cite{Skyrme:1961vq} arouses interesting, and the soliton picture consists with the QCD in the large $N_c$ limit~\cite{Witten:1983tw}. The $\Theta^+$, if exists, is the lightest member in the predicted antidecuplet~\cite{Manohar:1984ys,Chemtob:1985ar}. Its mass and width were predicted in the chiral soliton model~\cite{Jezabek:1987ns,Praszalowicz:2003ik,Walliser:1992vx,Diakonov:1997mm,Weigel:1998vt}.

In quark model language, the $\Theta^+$ is described as a pentaquark state $|uudd\bar{s}\rangle$. Unlike the other pentaquark $|qqqq\bar{q}\rangle$ states that the antiquark may have the same flavor with at least one quark, the lowest Fock state of the $\Theta^+$ composes of five quarks with the anti-quark being $\bar{s}$, which
is of different flavor from the other four $uudd$ quarks. Therefore it is easy to be identified from other ordinary baryons with minimal $|qqq\rangle$ configurations~\cite{Gao:1999ar}. For the pentaquark states that the antiquark has the same flavor of some quark, the mixing of the pentaquark state and the corresponding three-quark state and hybrid state makes the situation complicated, because any three-quark baryon may have five-quark components from both perturbative and nonperturbative aspects, such as the baryon-meson fluctuation picture~\cite{Brodsky:1996hc} and the light-cone Fock state expansion~\cite{Brodsky:1997de}. Since the $\Theta^+$ has the same quark constituents as the combination of $nK^+$ and $pK^0$, these two modes are expected as the primary decay channel, and thus are usually used in reconstructions in the experiments.

After the first report for the $\Theta^+$, the signals were observed by many groups~\cite{Nakano:2003qx,Barmin:2003vv,Aleev:2004sa,Airapetian:2003ri,Barth:2003es,%
Asratyan:2003cb,Asratian:2005rt,Chekanov:2004kn,Troyan:2004wp,Aslanyan:2004gs,Miwa:2006if,Panzarasa:2006ag}, and some groups confirmed their results with new data~\cite{Nakano:2008ee,Barmin:2006we,Barmin:2009cz,Barmin:2013lva,Aleev:2005yq}. All these results are briefly listed in Table \ref{thetay}. Negative results were also reported by many groups, and some early positive results in the photoproduction experiment by CLAS~\cite{Stepanyan:2003qr} and in the proton-proton collision by COSY-TOF~\cite{AbdelBary:2004ts} were rejected by their high-statistics experiments later~\cite{McKinnon:2006zv,AbdelBary:2006wd}. Those are discussed in the next section.
\begin{table}
\caption{Experiments with positive signals for the $\Theta^+$.}
\begin{ruledtabular}
\begin{tabular}{lccccccr}\label{thetay}
Group & Reaction & Mode & Signals & Mass (MeV) & Width (MeV) & Significance & Reference\\
\hline
LEPS & $\gamma\textrm{C}\rightarrow K^+K^-X$ & $K^+n$ & 19 & 1540 & $<25$ & 4.6$\sigma$ & \cite{Nakano:2003qx}\\
LEPS & $\gamma d\rightarrow K^+K^-np$ & $K^+n$ & 116 & 1524 & 12.7 & 5.1$\sigma$ & \cite{Nakano:2008ee}\\
DIANA & $K^+\textrm{Xe}\rightarrow K^0p\textrm{Xe}'$ & $K^0p$ & 29 & 1539 & $<9$ & 4.4$\sigma$ & \cite{Barmin:2003vv}\\
DIANA & $K^+\textrm{Xe}\rightarrow K_S^0pX$ & $K_S^0p$ & 85 & 1538 & 0.36 & 6.3$\sigma$ & \cite{Barmin:2006we,Barmin:2009cz,Barmin:2013lva}\\
CLAS & $\gamma p->\pi^+K^+K^-n$ & $K^+n$ & 41 & 1555 & $<26$ & 7.8$\sigma$ & \cite{Kubarovsky:2003fi}\\
SVD & $pA\rightarrow pK_S^0X$ & $K_S^0p$ & 50 & 1526 & $<24$ & 5.6$\sigma$ & \cite{Aleev:2004sa}\\
SVD & $pA\rightarrow pK_S^0X$ & $K_S^0p$ & 118, 187 & 1522, 1523.6 & $<14$ & 9.2, 6.0$\sigma$ & \cite{Aleev:2005yq}\\
HERMES & $\gamma d\rightarrow pK_S^0X$ & $K_S^0p$ & 52-74 & 1528 & 17 & 4.2-6.3$\sigma$ & \cite{Airapetian:2003ri}\\
SAPHIR & $\gamma p\rightarrow nK^+K_S^0$ & $K^+n$ & 55 & 1540 & $<25$ & 4.8$\sigma$ & \cite{Barth:2003es}\\
ITEP & $\nu_\mu(\bar{\nu}_\mu) A\rightarrow K_S^0pX$ & $K_S^0p$ & 32 & 1532.2 & $<12$ & 7.1$\sigma$ & \cite{Asratyan:2003cb,Asratian:2005rt}\\
ZEUS & $e^\pm p\rightarrow e^\pm p(\bar{p})K_S^0X$ & $K_S^0p$ & 221 & 1521.5 & 8 & 3.9-4.6$\sigma$ & \cite{Chekanov:2004kn}\\
JINR & $np\rightarrow npK^+K^-$ & $K^+n$ & 27 & 1541 & 8 & 6.8$\sigma$ & \cite{Troyan:2004wp}\\
JINR & $p\textrm{C}_3\textrm{H}_8\rightarrow pK_S^0X$ & $K_S^0p$ & 88 & 1540 & 9.2 & 5.5$\sigma$ & \cite{Aslanyan:2004gs}\\
E522 & $\pi^-p\rightarrow K^-X$ & -- & 183 & 1530 & 9.8 & 2.7$\sigma$ & \cite{Miwa:2006if}\\
OBELIX & $\bar{p}\textrm{He}\rightarrow pK_S^0X$ & $K_S^0p$ & 46, 59 & 1560.0 & narrow & 2.7$\sigma$ & \cite{Panzarasa:2006ag}\\
\end{tabular}
\end{ruledtabular}
\end{table}

\subsection{LEPS}

The first signal for a narrow $S=+1$ pentaquark candidate $\Theta^+$, which originally named as $Z^+$, was observed at the laser-electron photon facility at SPring-8 (LEPS)~\cite{Nakano:2003qx}. In this experiment, photons with maximum energy of 2.4 GeV were produced by Compton backscattering of 351-nm Ar laser from 8 GeV electrons in the SPring-8 storage ring, and only photons with energies above 1.5 GeV were tagged. A 0.5-cm thick plastic scintillator (SC) with C:H$\approx$1:1 was used as a target, and was located 9.5 cm downstream from the 5 cm thick the liquid-hydrogen (LH$_2$) target. Contributions from protons and neutrons were distinguished by the comparison between the events from the LH$_2$ and the SC selected under the same conditions.

The events with a $K^+K^-$ pair and the photon energy below 2.35 GeV were selected to reduce the contributions from nonresonant $K^+K^-$ productions. The missing-mass $MM_{\gamma K^+K^-}$ of the reaction $N(\gamma,K^+K^-)X$ was calculated with the assumption that the nucleon has the mass of 0.9389GeV/$c^2$ and zero momentum, and therefore the events were selected with the requirement $0.90<MM_{\gamma K^+K^-}<0.98\mathrm{GeV}/c^2$. The events with the invariant mass of the $K^+K^-$ pair from 1.00 to 1.04 GeV/$c^2$ were removed to eliminate the photoproduction of the $\phi$ meson. The events with the nucleon momentum smaller than 0.35 GeV/$c$ were rejected due to the large uncertainty of the calculated direction. Then a total of 109 events were selected as the signal sample.

The Fermi motion was taken into account via the process $\gamma n \rightarrow K^+\Sigma^-\rightarrow K^+\pi^-n$, and the corrected missing mass was $MM^c_{\gamma K^\pm}=MM_{\gamma K^\pm}-MM_{\gamma K^+K^-}+M_N$. A clear peak due to the reaction $\gamma p\rightarrow K^+\Lambda(1520)\rightarrow K^+K^-p$ was observed in the corrected $K^+$ missing-mass distribution within the 108 rejected events. In the corrected $K^-$ missing-mass distribution of the signal sample, a prominent peak at 1.54 GeV/$c^2$ is observed with 36 events in the peak region $1.51\leq MM^c_{\gamma K^-}<1.57\textrm{GeV}/c^2$. The number of events in the peak region above the background level is $19.0\pm2.8$ corresponding to a significance of $4.6\sigma$. The best fit of the mass is $1.54\pm0.01\textrm{GeV}/c^2$ with an estimated systematic error of 5 MeV/$c^2$. The upper limit of the width is 25MeV/$c^2$ with 90\% confidence level (C.L.).

A recent experiment via $\gamma d\rightarrow K^+K^-pn$ process by LEPS Collaboration reproduced this resonance~\cite{Nakano:2008ee}. This analysis based on the data collected in 2002-2003 with the statistics improved by a factor of eight. The photons, produced via the same approach as in the previous experiment, were alternatively injected into the liquid deuterium (LD$_2$) and the LH$_2$ targets. The events with a $K^+K^-$ pair were selected, and the vertex point of the kaon tracks was required to be within the target volume. The events with invariant mass of the $K^+K^-$ pair from 1.01 to 1.03 GeV/$c^2$ were removed to eliminate the contribution from the $\phi$ meson.

In this analysis, a minimum momentum spectator approximation (MMSA), that the spectator nucleon was assumed to have the minimum momentum $p_{\textrm{min}}$ for the given energy momentum of a $pn$ pair, was applied to improve the mass resolution. The $|p_\textrm{min}|$ was required to be less than 0.1GeV/$c$ and the effective photon energy $E^\textrm{eff}_\gamma=(M^{2}_{nK^+K^-}-M_N^2)/2M_N$ was required to be from 2.0 to 2.5 GeV in order to reduce the background from nonquasifree processes. For the LH$_2$ target, the requirements were $2.0<E_\gamma<2.4$ GeV and $0.9<MM_{\gamma K^+K^-}<0.98$ GeV/$c^2$ instead. In addition, a randomized minimum momentum (RMM) method was developed to estimate the reasonable spectrum shape.

The best fit peak position is $1524\pm2\pm3$ MeV/$c^2$, and the signal yield is estimated to be 116$\pm$21 events. The width for the best fit is 12.7$\pm$2.8 MeV/$c^2$. The smallest significance among the fit results with various background models is $5.1\sigma$. The differential cross section is estimated to be 12$\pm$2 nb/sr by assuming the isotropic production. The production ratio of $\Theta^+$ to $\Lambda(1520)$, by considering the partial decay branching ratios and the acceptance difference, is estimated to be 0.16$\pm$0.03.

\subsection{DIANA}

The DIANA Collaboration, by investigating the low-energy $K^+$Xe collisions, reported a resonant enhancement of $K^0p$ effective mass spectrum as an indication for the pentaquark candidate $\Theta^+$ baryon~\cite{Barmin:2003vv}. In this experiment, a separated $K^+$ beam with the momentum of 850 MeV/$c$ from the 10 GeV proton synchrotron at the ITEP was injected into the bubble chamber DIANA filled with liquid xenon. The momentum of an interacting $K^+$, from 750 to 0 MeV/$c$, was determined from the longitudinal coordinate of the interaction vertex with respect to the central position of the observed maximum due to decays of stopping $K^+$ mesons.

A total of 6334 events of the charge-exchange reaction $K^+\textrm{Xe}\rightarrow K^0X$ were selected. The $K_S^0$ was detected via the decay to the $\pi^0\pi^0$ or $\pi^+\pi^-$ and the $K_L^0$ was inferred from the non-observation of strange particles. The momenta of the proton and the $K_S^0$ were required to be above 180 and 170 MeV/$c$ respectively, and the $K^+$ range before the interaction $L_{K^+}$ was required to exceed 550 mm corresponding to the momentum less than 530 MeV/$c$. After these cuts 1112 events of the charge-exchange reaction $K^+\textrm{Xe}\rightarrow K^0p\textrm{Xe}'$ were selected.

The background of the $K^0p$ mass spectrum was fitted to a distribution obtained by the method of random stars and a simulation of non-resonant charge-exchange reaction by considering the momentum distribution of interacting $K^+$, the Fermi motion and binding energy of the target nucleon, the differential cross section for $K^0$ emission and actual conditions of the experiment. Then the 107 events from 1535 to 1545 MeV/$c^2$ with the estimated background of 83 events resulted in a statistical significance of $2.6\sigma$. If additional selections, $\theta_p<100\degree$ and $\theta_K<100\degree$ with respect to the $K^+$ direction in the laboratory frame and $\cos\Phi_{pK}<0$ for the azimuthal angle between the proton and the $K^0$ directions, were required in order to remove the events worst affected by rescatterings, the left 541 events resulted that 73 events were in the interval from 1535 to 1545 MeV/$c^2$ with the estimated background of 44 events. Then the fitted mass is 1539$\pm$2 MeV/$c^2$, and the width is no more than 9 MeV/$c^2$. The statistical significance of the signal is $4.4\sigma$.

The DIANA Collaboration, by including the new data, reported the updated result which was in agreement with the previous result~\cite{Barmin:2006we}. The events with a single proton and a $K^0_S\rightarrow\pi^+\pi^-$ were selected. The same cuts for the momenta of the proton and the $K^0_S$ were applied. A total of 2131 events were selected as the candidates for the $K^+\textrm{Xe}\rightarrow K^0p\textrm{Xe}'$ reaction. In the $M_{pK_S^0}$ distribution with the selections that $445<p_{K^+}<525$ MeV/$c$, $\theta_p<100\degree$ and $\theta_K<100\degree$, a peak at 1537$\pm$2 MeV/$c^2$ was observed. In the mass interval from 1532 to 1544 MeV/$c^2$, the statistical significance of the signal is 7.3, 5.3 and 4.3$\sigma$ deviations for the estimators $S/\sqrt{B}$, $S/\sqrt{S+B}$ and $S/\sqrt{S+2B}$, where the signal $S=60$ events and the background $B=68$ events. The intrinsic width $\Gamma$ of the resonance is estimated to be 0.36$\pm$0.11 MeV/$c^2$ from the formula
\begin{equation}\label{iwidth}
\Gamma=\frac{N^\textrm{peak}}{N^\textrm{bkgd}}\frac{\sigma^\textrm{CE}}{107\textrm{mb}}\frac{\Delta m}{B_iB_f},
\end{equation}
where $N^\textrm{peak}$ and $N^\textrm{bkgd}$ are the numbers of events in the peak and background, $\sigma^\textrm{CE}$ is the cross section for $K^+n\rightarrow K^0p$, $\Delta m$ is the $M_{pK^0_S}$ interval under the peak and $B_i=1/2$ and $B_f=1/2$ are branching fractions for the initial and final states.

With the statistics of the charge-exchange reaction increased by 11\% as compared to~\cite{Barmin:2006we}, the DIANA Collaboration reanalyzed the data~\cite{Barmin:2009cz}. The events with a single proton and a $K^0_S\rightarrow\pi^+\pi^-$ were selected, and the momenta of the proton and the $K^0_S$ were required to be above 170 MeV/$c$ and 160 MeV/$c$ respectively. Another selection $\tau<3\tau_0$, where $\tau$ is the $K_S^0$ measured proper lifetime and $\tau_0$ is the mean value, was appied to further reject the events that the $K_S^0$ scattered by small angles in the liquid xenon but passed the pointback criteria. As suggested by the simulation, the selection that $|A_m|=|M_{pK_S^0}-\sqrt{s}|/(M_{pK_S^0}+\sqrt{s})<0.015$, where $\sqrt{s}$ is the center-of -mass energy of the $K^+n$ with a stationary free neutron assumption, was used to suppress rescatterings. Then the fitted mass is 1538$\pm$2 MeV/$c^2$ with 99.9$\pm$18.2 signal events corresponding to a statistical significance of $8\sigma$ estimated as $S/\sqrt{B}$ or $6\sigma$ estimated as $S/\sqrt{S+B}$. If an additional selection $\theta_p<100\degree$ and $\theta_K<100\degree$, or longitudinal momentum $p_L(pK_S^0)>80\textrm{MeV}/c$, or the cosine of the $K_S^0$ emission angle in the $pK_S^0$ rest frame $\cos\theta^*>0.2$ was applied, the number of the signal events was 89.1$\pm$15.4, 103.1$\pm$16.8 or 97.0$\pm$16.3 respectively. The intrinsic width of the $\Theta^+$ estimated from the formula (\ref{iwidth}) is 0.39$\pm$0.10 MeV/$c^2$. In the recent reanalysis by the DIANA Collaboration~\cite{Barmin:2013lva}, the events that passed the selections, $L_{K^+}>520$ mm, $\tau<3\tau_0$, $p_{K_S^0}>155$ MeV/$c$, $p_p>165$ MeV/$c$ and the angle between the pions $\theta_{\pi\pi}<150\degree$, were selected. An enhancement at $M_{pK_S^0}=1538$ MeV/$c^2$ was observed. The $M_{pK_S^0}$-$p_{K^+}$ plots with additional selection $\theta_p,\theta_K<100\degree$ or $p_L(pK_S^0)>120$ MeV/$c$ suggests a limited preferred range of the $p_{K^+}$. The statistical significance estimated as the log-likelihood difference between the signal and null hypotheses reaches 6.0$\sigma$ with $77.92\pm16.6$ signals and 6.3$\sigma$ with $84.82\pm16.98$ signals for the two additional selections respectively. In the $pK_S^0$ effective mass distributions with common selections $p_T(pK_S^0)<300$ MeV/$c$ and $445<p_{K^+}<535$ MeV/$c$ added, the number of the signals is $67.49\pm14.38$ and $68.09\pm15.14$ respectively, corresponding to the significance of 5.9$\sigma$ and 5.6$\sigma$. The intrinsic width estimated from (\ref{iwidth}) is $0.36\pm0.11$ MeV/$c^2$

\subsection{CLAS}

The CLAS Collaboration reported the observation of the $\Theta^{+}$ in the photoproduction reaction from the proton target~\cite{Kubarovsky:2003fi}. The data from three distinct runs with the photon beam energy ranges of 3.2-3.95, 3-3.25 and 4.8-5.47 GeV were included, and two reactions $\gamma p\rightarrow\pi^+K^+K^-n$ and $\gamma p\rightarrow K^+K^-p$ were analyzed.

In the analysis, the events with a $\pi^+$, $K^+$ and $K^-$ identified by the time of flight in the final state were selected for the reaction $\gamma p\rightarrow\pi^+K^+K^-n$. The neutron peak were clearly seen in the missing mass distributions for the reaction $\gamma p\rightarrow\pi^+K^+K^-X$. A total of 14 000 events within $2\sigma$ of the neutron peak were retained. Then about 200 $\phi$ mesons were removed by the requirement $M_{K^+K^-}>1.06$ GeV/$c^2$. No obvious structure was observed in the $nK^+$ invariant mass spectrum from the missing mass for the reaction $\gamma p\rightarrow \pi^+K^-X$. However, a clear peak appeared after the cut $\cos\theta^*_{\pi^+}>0.8$ applied, where the $\theta^*$ was the center-of-mass angle in relative to the direction of the photon beam. By varying the angular cut from 0.7 to 0.9, the peak remained clearly visible. In addition, another angular cut $\cos\theta^*_{K^+}>0.6$ was applied to suppress the background from the meson and baryon decays.

Fitted by the sum of a Gaussian function and a background function obtained from the simulation, a narrow resonance at $M=1555\pm1\pm10$ MeV/$c^2$ with the width $\Gamma<26$ MeV/$c^2$ was observed in the $nK^+$ invariant mass spectrum. The number of the signal is 41$\pm$10, corresponding to a statistical significan
ce of 7.8$\pm$1.0 $\sigma$ estimated as $S/\sqrt{B}$.

In the later high-statistics photoproduction reactions from the proton target by CLAS~\cite{Battaglieri:2005er,DeVita:2006ha}, this resonance was not reproduced, but the final states in those analyses were different. Besides, a resonance as the candidate for the $\Theta^+$ was also reported in the early $\gamma d$ experiment by CLAS~\cite{Stepanyan:2003qr}, but the result was rejected by the later high-statistics $\gamma d$ experiment~\cite{McKinnon:2006zv}.

\subsection{SVD}

A resonance of $pK^0_S$ invariant mass spectrum in the reaction $pN\rightarrow pK_S^0X$ was observed in the data from the SVD-2 experiment~\cite{Aleev:2004sa}. In this experiment, the proton beam with the momentum of 70 GeV/$c$ from the ITEP accelerator was injected into an active target which consisted of five microstrip silicon detectors of 300 $\mu$m, a lead foil of 220 $\mu$m and a carbon target of 500 $\mu$m.

In the analysis, the events with the multiplicity of charged particles in the primary vertex not more than five were selected in order to suppress the combinatorial background and to reduce the probability of events involving rescatterings and the background from the $K_S^0$ produced in the central rapidity region. The $K_S^0$ mesons were reconstructed by their decay to the $\pi^+\pi^-$ mesons. The events with $M_{p\pi^-}<1.12$ GeV/$c^2$ were rejected to eliminate the background from $\Lambda$ decays. Then about 3800 $K_S^0$ mesons with the decay length not larger than 35 mm were selected for further analysis. The momentum of the proton was required to be from 4 to 21 GeV/$c$. Two additional cuts, $490\leq M_{\pi^+\pi^-}\leq 505\textrm{MeV}/c^2$ and $\cos\alpha\geq0$ where $\alpha$ is the $pK_S^0$ emission angle in the center of mass system of $pN$, were applied to improve the resolution of $M_{pK_S^0}$ spectrum and to suppress the background from the $\pi$ mesons misidentified as protons. In the mass region above 1550 MeV/$c^2$, the selection $p_{K_S^0}\leq p_p$ was used to suppress the background from the $\Sigma^{*+}$ resonances.

In the region $1.3<M_{pK_S^0}<1.7$ GeV/$c^2$, a resonance of $M=1526\pm3\pm3$ MeV/$c^2$ and $\Gamma<24$ MeV/$c^2$ is observed. Within the mass window $\Delta M=45$ MeV/$c^2$, the signal number estimated by the simulation is 50 above the background of 78 events, corresponding to a statistical significance of $5.6\sigma$ estimated as $S/\sqrt{B}$. The total cross section for $\Theta^+$ production in proton-nucleus interactions is estimated to be 30-120 $\mu$b with the detection efficiency of 0.07\% for the $\Theta^+$ baryon.

The SVD Collaboration updated their results by analyzing two different samples of $K_S^0$ events~\cite{Aleev:2005yq}. For the first analysis, the events with the multiplicity not more than seven and the $K_S^0$ decay length less than 35 mm were selected. In the $pK_S^0$ invariant mass spectrum, a clear excess at $M=1522\pm3$ MeV/$c^2$ with 205 signal over 1050 events is observed. With two additional cuts, $3<p_p<10\textrm{GeV}/c$ and the number of tracks associated with the primary vertex $N_\textrm{tracks}>4$, applied, there are 118 signals over 162 background events in the window from 1500 to 1540 MeV/$c^2$, corresponding to a statistical significance of 9.2$\sigma$ estimated as $S/\sqrt{B}$. The upper limit on the intrinsic width is estimated as 14 MeV/$c^2$ with 95\% C.L.. For the second analysis, the events with the multiplicity not more than six and the $K_S^0$ decay length from 35 to 600 mm were selected. The candidates with $M_{p\pi^-}<1.12$ GeV/$c^2$ were rejected, and a selection on the proton momentum $8<p_p<15\textrm{GeV}/c$ was applied. In the $pK_S^0$ invariant mass spectrum, an enhancement at $M=1523.6\pm3.1$ with 187 signals over 940 background events is observed, corresponding to a statistical significance of $6.0\sigma$ estimated as $S/\sqrt{B}$. This resonance was verified not to be a reflection from other resonances.

\subsection{HERMES}

A signal for a exotic baryon state as $\Theta^+$ candidate was reported by the HERMES Collaboration in the quasi-real photoproduction on the deuterium gas target through the $pK_S^0\rightarrow p\pi^+\pi^-$ channel~\cite{Airapetian:2003ri}. The data were obtained with the 27.6 GeV positron beam of the HERA storage ring at DESY.

The events with a $\pi^+\pi^-$ pair and a coincident proton tracks were selected, and the momenta of the protons and the pions were restricted to 4-9 GeV/$c$ and 1-15 GeV/$c$ respectively. Due to the intrinsic tracking resolution, the selections that the minimum distance of approach between two pion tracks was less tan 1 cm, the minimum distance of approach between the proton and the reconstructed $K_S^0$ tracks was less than 6 mm, the radial distance of the production vertex from the positron beam axis was less than 4 mm, the $z$ coordinate of the vertex along the beam direction was $-18<z<18$ cm within the $\pm20$ cm long target and the $K_S^0$ decay length was greater than 7 cm, were required. To suppress the background from $\Lambda$ decays, the events with $M_{p\pi^-}$ within $M_\Lambda\pm2\sigma$ observed in this experiment were rejected. In addition, the $M_{\pi^+\pi^-}$ was required to be within $\pm2\sigma$ of the $K_S^0$ peak.

In the analysis for the $\Theta^+$, two different fits were applied with the background estimated in three different models, and the results consisted with each other. Among these results, the number of signals is 52-74 over the background of 144-158 events. The mass of the resonance is $M=1528\pm2.6\pm2.1$ MeV/$c^2$, and the extracted intrinsic width is $\Gamma=17\pm9\pm3$ MeV/$c^2$. The statistical significance is 4.2-6.3$\sigma$ estimated as $S/\sqrt{B}$ and 3.4-4.3$\sigma$ estimated as $S/\delta S$, where the $\delta S$ is its fully correlated uncertainty. By taking the acceptance and the branching ratio to $pK_S^0$ as 0.05\% and 25\%, the photoproduction cross section was estimated to be from 100 to 220 nb $\pm$ 25\%, corresponding to the ratio of the $\Theta^+$ cross section to that of the $\Lambda(1520)$ between 1.6 and 3.5. Besides, the observed $\Theta^+$ is likely to be an isoscalar due to the absence of a resonance in $pK^+$ invariant mass spectrum.

\subsection{SAPHIR}

 The resonance as $\Theta^+$ candidate was reported by the SAPHIR Collaboration through the photoproduction reaction $\gamma p\rightarrow nK^+K_S^0$~\cite{Barth:2003es}. The data were taken in 1997 and 1998 with a trigger requiring at least two charged particles and the SAPHIR detector at the Bonn Electron Stretcher Accelerator (ELSA). The photons were produced from the ELSA electron beam via bremsstrahlung in a copper foil radiator. The minimum momenta of the detected charged pions and the protons were 50 and 150 GeV/$c$.

In this experiment, the $\Theta^+$ resonance was analyzed through the reactions $\gamma p\rightarrow \Theta^+ K_S^0$, $\Theta^+\rightarrow nK^+$ and $K_S^0\rightarrow \pi^+\pi^-$. The events that passed any of the hypotheses, $\gamma p\rightarrow p\pi^+\pi^-$, $\gamma p\rightarrow pK^+K^-$, $\gamma p\rightarrow pe^+e^-$, $\gamma p\rightarrow p\pi^+\pi^-\pi^0$, $\gamma p\rightarrow p\pi^-K^+$ and $\gamma p\rightarrow p\pi^-\pi^0K^+$, were rejected. Two cuts $480<M_{\pi^+\pi^-}<518$ MeV/$c^2$ and $\cos\theta_{K_S^0}>0.5$, where the $\theta_{K_S^0}$ was the $K_S^0$ production angle in the center-of-mass frame, were applied. The contamination of $\Lambda(1520)$ in the $nK^+$ invariant mass spectrum was reduced by the second cut.

A clear signal in the $nK^+$ invariant spectrum is observed at $M=1540\pm4\pm2$ MeV/$c^2$ with $63\pm13$ signal events, corresponding to a statistical significance of $4.8\sigma$ estimated as $S/\sqrt{S+B}$. By counting the four bins around the $\Theta^+$ center, the number of the signal is 55 over a background of 56 events. The upper limit on the width is 25 MeV with 90\% C.L.. The photoproduction cross section for the $\Theta^+$ is estimated to be about 300 nb from the number of events observed in the full angular range. In addition, due to the absence of a signal at expected strength in the $pK^+$ invariant mass spectrum in the reaction $\gamma p\rightarrow pK^+K^-$, the observed $\Theta^+$ is likely to be an isoscalar.

\subsection{ITEP}

Several resonances in the $pK_S^0$ invariant mass spectrum were observed from the old neutrino bubble chambers data by the ITEP scientists~\cite{Asratyan:2003cb,Asratian:2005rt}. The database comprised 120 000 $\nu_\mu$ and $\bar{\nu}_\mu$ induced charge-current (CC) events collected both in the BEBC experiments WA21 filled with hydrogen, WA25 filled with deuterium and WA59 filled with neon-hydrogen mix at CERN and in the experiments E180 and E632 filled with neon-hydrogen mix at FermiLab.

In the analysis, any additional proton was required to be soft with the momentum less than 400 MeV/$c$. The longitudinal momentum of the $pK_S^0$ system was required to be less than 2.5 GeV/$c$. The background of the $M_{pK_S^0}$ distribution was estimated by randomly pairing a $K_S^0$ and a proton from two events in the same $\nu_\mu$ or $\bar{\nu}_\mu$ subsample.

With the deuterium and neon data combined, the masses of the three resonances observed in $M_{pK_S^0}$ spectrum are $1532.2\pm1.3$ MeV/$c^2$ with $32\pm7$ signals, $1577.7\pm1.9$ MeV/$c^2$ with $31\pm7$ signals and $1658.6\pm4.4$ MeV/$c^2$ with $33\pm10$ signals, and the significance levels are $7.1\sigma$, $5.0\sigma$ and $4.5\sigma$ respectively. The observed widths of the 1532 and 1578 MeV/$c^2$ resonances are restricted to be less than 12 and 23 MeV/$c^2$ with 90\% C.L.. However, no signals were detected in the free protons data.

\subsection{ZEUS}

A resonance in $K_S^0p$ and $K_S^0\bar{p}$ invariant mass spectrum was observed by the ZEUS Collaboration in the deep inelastic scattering (DIS) experiment~\cite{Chekanov:2004kn}. The data were collected with the ZEUS detector at HERA from 1996 to 2000. The center-of-mass energy was 300 GeV and 318 GeV for $e^+p$ scatterings and was 318 GeV for $e^-p$ scatterings.

The events were selected with the requirements that the exchanged photon virtuality $Q^2\geq1$ GeV$^2$, the corrected energy of the scattered electron or positron $E_{e'}\geq8.5$ GeV, $35\leq\delta=\sum E_i(1-\cos\theta_i)\leq60$ GeV where $E_i$ is the energy of the $i$-th calorimeter cell and $\theta_i$ is the polar angle, the Bjorken scaling variable $y$ reconstructed using the electron method $y_e\leq0.95$ and that reconstructed using the Jacquet-Blondel (JB) method $y_\textrm{JB}\geq0.01$, and the vertex position $|z|\leq 50$ cm. The tracks were required to pass through not less than five central tracking detector (CTD) superlayers and to have transverse momenta larger than 0.15 GeV/$c$ and pseudorapidity in the laboratory frame $|\eta|\leq1.75$.

The candidates for $K_S^0$ mesons were selected through the decay $K_S^0\rightarrow\pi^+\pi^-$ with the cuts,  $M_{e^+e^-}\geq50$ MeV/$c^2$ to eliminate tracks from photon conversions, $M_{p\pi}\geq1121$ MeV/$c^2$ to eliminate $\Lambda$ and $\overline{\Lambda}$ contaminations, $483\leq M_{\pi^+\pi^-}\leq513$ MeV/$c^2$, $p_T(K_S^0)\geq0.3$ GeV/$c$ and $|\eta(K_S^0)|\leq1.5$. The candidates for the protons or the antiprotons were selected through the energy-loss measurement in the CTD, $dE/dx$, with the cuts, $dE/dx>1.15$ mips and $f\leq dE/dx\leq F$ where $ f=0.35/p^2+0.8$ and $F=1.0/p^2+1.2$ for positive tracks and $f=0.3/p^2+0.8$ and $F=0.75/p^2+1.2$ for negative tracks where $p$ is the total track momentum in GeV. The momentum of the proton or the antiproton was required to below 1.5 GeV/$c$.

For the analysis of the $K_S^0p(\bar{p})$ invariant mass spectrum, the $Q^2$ was integrated above various minimum values from 1 to 50 GeV$^2$, and a peak was seen around 1520 MeV/$c^2$ for $Q^2\geq10$ GeV$^2$. From a fit in the region $Q^2\geq20$ GeV$^2$, the peak is at $M=1521.5\pm1.5^{+2.8}_{-1.7}$ MeV/$c^2$ with the measured Gaussian width of $\sigma=6.1\pm1.6^{+2.0}_{-1.4}$ MeV/$c^2$ and the extracted intrinsic width $\Gamma=8\pm4$ MeV/$c^2$. The number of signal events is $221\pm48$, corresponding to a statistical significance of 3.9-4.6$\sigma$. For the $K_S^0\bar{p}$ channel only, the signal number is $96\pm34$. It is the first signal for the antiparticle of the $\Theta^+$.

\subsection{JINR}

Several resonances in the $nK^+$ invariant mass spectrum were reported in a neutron experiment at Joint Institute for Nuclear Research (JINR)~\cite{Troyan:2004wp} through the reaction $np\rightarrow npK^+K^-$. The data were obtained in an exposure of a 1-m H$_2$ bubble chamber of the Vekster and Baldin Laboratory of High Energies (LHE) at JINR to a quasimonochromatic neutron beam. The momenta of the neutrons of the selected events were $p_n=5.20\pm0.12\textrm{GeV}/c$.

In the analysis, the background curves were taken in the Legendre polynomial form and the resonance curves were taken in the Breit-Wigner form. A resonance at $M=1541$ MeV/$c^2$ as well as other resonances was observed in $M_{nK^+}$ spectrum with either the sample of 1558 events under the condition $|\cos\theta_n^*|>0.85$ or the sample of 1157 events under the condition $|\cos\theta_n^*|<0.85$, where $\theta_n^*$ was the emission angle of the neutron in the center-of-mass frame. With additional selections, the resonance at $M=1541\pm4$ MeV/$c^2$ with $\Gamma=8\pm4$ MeV/$c^2$ is of $6.8\sigma$ statistical significance under the cuts $\cos\theta_n^*\geq-0.70$ and $2.05\leq M_{nK^+K^-}\leq2.15$ GeV/$c^2$ or $2.24\leq M_{nK^+K^-}\leq2.28$ GeV/$c^2$, the resonance at $M=1606\pm5$ MeV/$c^2$ with $\Gamma=11\pm6$ MeV/$c^2$ is of $5.2\sigma$ statistical significance under the cuts $-0.70\leq\cos\theta_n^*\leq0.70$ and $2.10\leq M_{nK^+K^-}\leq2.24$ GeV/$c^2$ or $2.28\leq M_{nK^+K^-}\leq2.50$ GeV/$c^2$ and the resonance at $M=1687\pm7$ MeV/$c^2$ with $\Gamma=24\pm8$ MeV/$c^2$ is of $6.8\sigma$ statistical significance under the cut $2.24\leq M_{nK^+K^-}\leq2.29$ GeV/$c^2$. Including the other resonances of lower statistical significance, the spins of these resonances are analyzed to be $5/2$, $7/2$ and $9/2$ respectively using the formula for the rotational bands $M_J=M_0+J(J+1)/2I$, where the $M_0$ and $I$ are the mass and the inertia moment of the soliton.

More than one $S=+1$ pentaquark-like narrow resonances were also reported in $pK_S^0$ invariant mass spectrum via the reaction $p\textrm{C}_3\textrm{H}_8\rightarrow pK_S^0X$~\cite{Aslanyan:2004gs}. This investigation was performed at LHE-JINR. The data were obtained in an exposure of a 2-m C$_3$H$_8$ bubble chamber to the proton beam with the momentum of 10 GeV/$c$. The $pK_S^0$ effective mass distribution was analyzed in different regions, $0.350\leq p_p\leq0.900$ GeV/$c$, $0.9\leq p_p\leq1.7$ GeV/$c$ or $p_p\geq1.7$ GeV/$c$. Based on the data in the region $0.350\leq p_p\leq0.900$ GeV/$c$, the fitted masses and widths of the resonances are $M=1540\pm8$ MeV/$c^2$ with $\Gamma=9.2\pm1.8$ MeV/$c^2$, $M=1613\pm10$ MeV/$c^2$ with $\Gamma=16.1\pm4.1$ MeV/$c^2$ and $M=1821\pm11$ MeV/$c^2$ with $\Gamma=28.0\pm9.4$ MeV/$c^2$. The statistical significance of these resonances is estimated as $5.5\pm0.5\sigma$, $4.8\pm0.5\sigma$ and $5.0\pm0.6\sigma$ based on the data with identified protons and positively charged tracks $p_p\geq1.7\textrm{GeV}/c$. The total cross section for the $\Theta^+$ production in $p$C$_3$H$_8$ interactions is estimated to be about 90 $\mu$b.

\subsection{E522}

The E522 experiment at the K2 beam line of the KEK 12 GeV proton synchrotron was performed to search for the pentaquark candidate $\Theta^+$ via the reaction $\pi^-p\rightarrow K^-X$~\cite{Miwa:2006if}. In this experiment, a scintillation fiber ((CH)$_n$) and a bulk of polyethylene ((CH$_2$)$_n$) were used as the targets. The $\pi^-$ beam of 1.87 GeV/$c$ was irradiated to both the targets, and the $\pi^-$ beam of 1.92 GeV/$c$ was irradiated to the polyethylene target only. In order to estimate the contribution from the carbon nucleus, the data with a carbon target were taken also.

The $\Theta^+$ was analyzed from the missing mass spectrum of the reaction $\pi^-p\rightarrow K^-X$. The incident $\pi^-$ was identified using the time-of-flight information between T1 and T2 within $\pm3\sigma$ region of the time resolution of 70 ps. The $\pi^-$ momentum was cut at 95\% C.L., and it was required to be in the interval from 1.8 to 2.0 GeV/$c$. The scattered $K^-$ mesons were selected by the momentum dependent mass cut within $3\sigma$ region. The momentum was analyzed using the Runge-Kutta method, and the region $\chi^2<6.0$ was selected. The vertex point, which was calculated by the closest distant point between tracks of beam and the outgoing particles, was required to be less than 80 mm from the target center. The closest distance between the tracks from the beam and the outgoing particles at the vertex was required to be less than 7 mm in order to eliminate the rescatterings.

A peak is observed at $M=1530^{+2.2}_{-1.9}\ ^{+1.9}_{-1.3}$ MeV/$c^2$ with $\Gamma=9.8^{+7.1}_{-3.4}$ MeV/$c^2$ in the missing mass spectrum of 1.92 GeV/$c$ beam data, while no peak is observed in the missing mass spectrum of 1.87 GeV/$c$ beam data. Fitted with the fixed width of 13.4 MeV/$c^2$, the number of the signal events is $183\pm71\pm10$, corresponding to a statistical significance of 2.7$\sigma$ estimated as $S/\sqrt{S+B}$ and $2.5\sigma$ estimated as $S/\sqrt{\delta S}$. This bump was not claimed as an evidence of $\Theta^+$ due to the low statistical significance. The upper limit on the $\Theta^+$ production cross section in the reaction $\pi^-p\rightarrow K^-\Theta^+$ is estimated as 1.8 and 3.9 $\mu$b with the beam momentum of 1.87 and 1.92 GeV/$c$ respectively.

\subsection{OBELIX}

A resonance was observed as the $\Theta^+$ by reanalyzing the antiproton $^4$He annihilation data collected by the OBELIX Collaboration~\cite{Panzarasa:2006ag}. In the experiment, the antiproton beam with the momentum of 200 MeV/$c$ was extracted from the LEAR at CERN and slowed down in the $^4$He target. The $\Theta^+$ was searched for through the $pK_S^0$ decay mode in four channels
\begin{eqnarray}
\bar{p}^4\textrm{He}\rightarrow(p\pi^-\pi^-\pi^+\pi^+)nnX,\quad(p\pi^-\pi^+K^-)p_snX,\quad(p\pi^-\pi^+\pi^+K^-)nnX,\quad pp\pi^+\pi^-\pi^-)nX,\nonumber
\end{eqnarray}
where the $p_s$ is undetected slow proton with the momentum below 300 MeV/$c$ and the $X$ is the undetected neutral mesons like $\pi^0$ and $K_L^0$. The number of the selected events for each channel was 20 407, 702, 357 and 10 132 respectively.

In the analysis, the $K_S^0$ was identified via the $\pi^+\pi^-$ with the invariant mass within $495\pm20$ MeV/$c^2$. For the $p\pi^-\pi^-\pi^+\pi^+$ final state, the $\Theta^+$ might present only in the events with two $K_S^0$. Then two peaks were observed in the $M_{p\pi^+\pi^-}$ spectrum at 1500 and 1550 MeV/$c^2$, and the latter one was associated to the $\Theta^+$. For the other three final states, the similar structure was also observed in the $M_{p\pi^+\pi^-}$ spectrum but low statistics.

The fitted mass of the resonance is $1560.0\pm3.7$ MeV/$c^2$ with 46 or 59 signal events under the peak depending on the fitting ways. The Gaussian width is $10.76\pm4.51$ MeV/$c^2$ which is equal to the resolution 9.3 MeV/$c^2$, and therefore the intrinsic width is narrow. The statistical significance is $2.7\sigma$ estimated as $S/\sqrt{S+2B}$. It is not sufficient to be claimed as an evidence.

\section{Experiments with negative results for $\Theta^+$}

Although many experimental results supported the existence of the $\Theta^+$, null results were also reported by many groups especially in high energy experiments~\cite{Stepanyan:2003qr,McKinnon:2006zv,Niccolai:2006td,Kubarovsky:2003fi,Battaglieri:2005er,DeVita:2006ha,AbdelBary:2004ts,AbdelBary:2006wd,%
Link:2006yh,Samoylov:2006sb,Bai:2004gk,Aubert:2005qi,Wang:2005fc,Aubert:2007qea,Schael:2004nm,Abdallah:2007ad,Achard:2006be,Aktas:2006ic,%
Nekipelov:2006if,Alt:2003vb,Litvintsev:2004yw,Abt:2004tz,Antipov:2004jz,Pinkenburg:2004ux,Longo:2004gd,Adamovich:2005ns,Napolitano:2004mn,%
Miwa:2007xk,Shirotori:2012ka,Abe:2005gy} as briefly listed in the Table \ref{thetan}. Two collaborations, the CLAS and the COSY-TOF, who had observed the signals in their early experiments, failed to reproduce the $\Theta^+$ signals in their improved high statistics experiments~\cite{Stepanyan:2003qr,McKinnon:2006zv,Niccolai:2006td,Kubarovsky:2003fi,Battaglieri:2005er,DeVita:2006ha,AbdelBary:2004ts,AbdelBary:2006wd}. This makes the existence of the $\Theta^+$ a puzzle. Therefore the production mechanism and the analysis method should be investigated and reexamined.
\begin{table}
\caption{Experiments with negative results for the $\Theta^+$.}
\begin{ruledtabular}
\begin{tabular}{lccccr}\label{thetan}
Group & Reaction & Mode & Upper limit & Confidence & Reference\\
\hline
CLAS & $\gamma d\rightarrow pK^-K^+n$ & $K^+n$ & $\sigma<0.3$ nb & 95\% & \cite{Stepanyan:2003qr,McKinnon:2006zv}\\
    & $\gamma d\rightarrow \Lambda K^+n$ & $K^+n$ & $\sigma<5$ nb & 95\% & \cite{Niccolai:2006td}\\
    & $\gamma p\rightarrow \overline{K}^0K^+n$ & $K^+n$ & $\sigma<0.8$ nb & 95\%&       \\
    &                               &       & $N(\Theta^+)/N(\Lambda(1520))<0.22\%$ & 95\% & \cite{Battaglieri:2005er,DeVita:2006ha}\\
    & $\gamma p\rightarrow \overline{K}^0K^0p$ & $K_S^0p$ & $\sigma<1.5$ nb & 95\% &        \\
COSY-TOF & $pp\rightarrow \Sigma^+K^0p$ & $K_S^0p$ & $\sigma<0.15$ $\mu$b & 95\% & \cite{AbdelBary:2004ts,AbdelBary:2006wd}\\
FOCUS & $\gamma\textrm{BeO}\rightarrow pK_S^0X$ & $K_S^0p$ & $\sigma(\Theta^+)\mathcal{B}(pK_S^0)/\sigma(K(892)^+)<0.13\%$ & 95\% &      \\
    &       &       & $\sigma(\Theta^+)\mathcal{B}(pK_S^0)/\sigma(\Sigma(1385)^\pm)<2.3\%$ & 95\% & \cite{Link:2006yh}\\
NOMAD & $\nu_\mu A\rightarrow K_S^0pX$ & $K_S^0p$ & $N(\Theta^+)/N_\textrm{events}<2.13\times10^{-3}$ & 90\% & \cite{Samoylov:2006sb}\\
BES & $\psi(2S),J/\psi$ decays & $K^+n$, $K_S^0p$ & see Eq.(\ref{bes}) & 90\% & \cite{Bai:2004gk}\\
BaBar & $e^+e^-\rightarrow\Upsilon(4S)\rightarrow pK_S^0X$ & $K_S^0p$ & $N(\Theta^+)/N_\textrm{events}<1.8\times10^{-4}$ & 95\% &   \\
    & $e^+e^-\rightarrow q\bar{q}\rightarrow pK_S^0X$ & $K_S^0p$ & $N(\Theta^+)/N_\textrm{events}<5.0\times10^{-5}$ & 95\% & \cite{Aubert:2005qi}\\
    & $B^0\rightarrow p\bar{p}K_S^0$ & $K_S^0p$ & $\mathcal{B}(\Theta^+)\cdot\mathcal{B}(pK_S^0)<0.5\times10^{-7}$ & 95\% & \cite{Aubert:2007qea}\\
Belle & $B^0\rightarrow p\bar{p}K_S^0$ & $K_S^0p$ & $\mathcal{B}(\Theta^+)\cdot\mathcal{B}(pK_S^0)<2.3\times10^{-7}$ & 90\% & \cite{Wang:2005fc}\\
    & $KN\rightarrow pK_S^0X$ & $K_S^0p$ & $N(\Theta^+)/N(\Lambda(1520))<2.5\%$ & 90\% &    \\
    & $K^+n\rightarrow pK_S^0$ & $K_S^0p$ & $\Gamma<0.64$ MeV & 90\% & \cite{Abe:2005gy}\\
ALEPH & $Z\rightarrow pK_S^0X$ & $K_S^0p$ & $N(\Theta^+)/N_\textrm{events}<2.5\times10^{-3}$ & 95\% & \cite{Schael:2004nm}\\
DELPHI & $Z\rightarrow pK_S^0X$ & $K_S^0p$ & $N(\Theta^+)/N_\textrm{events}<2.0\times10^{-3}$ & 95\% & \cite{Abdallah:2007ad}\\
L3 & $\gamma\gamma\rightarrow p(\bar{p})K_S^0X$ & $K_S^0p$ & $N(\Theta^+)/N_\textrm{events}<4.7\times10^{-3}$ & 95\% & \cite{Achard:2006be}\\
H1 & $ep\rightarrow ep(\bar{p})K_S^0$ & $K_S^0p$ & $\sigma<120-360$ pb & 95\% & \cite{Aktas:2006ic}\\
COSY-J\"{u}lich & $pp\rightarrow pK^0\pi^+\Lambda$ & $K^0p$ & $\sigma<58$ nb & 95\% & \cite{Nekipelov:2006if}\\
NA49 & $pp\rightarrow pK_S^0X$ & $K_S^0p$ & not observed & -- & \cite{Alt:2003vb}\\
CDF & $p\bar{p}\rightarrow pK_S^0X$ & $K_S^0p$ & $N(\Theta^+)<89,76$ & 90\% & \cite{Litvintsev:2004yw}\\
HERA-B & $p\textrm{C}\rightarrow pK_S^0X$ & $K_S^0p$ & $N(\Theta^+)/N(\Lambda(1520))<2.7\%$ & 95\% & \cite{Abt:2004tz}\\
SPHINX & $pN\rightarrow nK^+K_S^0N$ & $K^+n$ & $\sigma<26$ nb & 90\% &  \\
    & $pN\rightarrow pK_S^0K_L^0N$ & $K_S^0p$ & $\sigma<42$ nb & 90\% &  \\
    & $pN\rightarrow pK_L^0K_S^0N$ & $K_L^0p$ & $\sigma<39$ nb & 90\% &  \\
    & $pN\rightarrow pK_S^0K_S^0N$ & $K_S^0p$ & $\sigma<52$ nb & 90\% & \cite{Antipov:2004jz}\\
PHENIX & $d\textrm{Au}\rightarrow K^-\bar{n}X$ & $K^-\bar{n}$ & not observed & -- & \cite{Pinkenburg:2004ux}\\
HyperCP & $p(\pi^+,K^+)\textrm{Cu}\rightarrow p(\bar{p})K_S^0X$ & $K_S^0p$ & $N(\Theta^+)/N_\textrm{events}<0.3\%$ & 90\% & \cite{Longo:2004gd}\\
LASS & $K^+p\rightarrow K^+n\pi^+$ & $K^+n$ & no narrow resonance & -- & \cite{Napolitano:2004mn}\\
WA89 & $\Sigma^-\textrm{C(Cu)}\rightarrow pK_S^0$ & $K_S^0p$ & $\sigma<7.2$ $\mu$b & 99\% & \cite{Adamovich:2005ns}\\
E559 & $K^+p\rightarrow \pi^+X$ & -- & $d\sigma/d\Omega<3.5$ $\mu$b/sr & 90\% & \cite{Miwa:2007xk}\\
J-PARC & $\pi^-p\rightarrow K^-X$ & -- & $d\sigma/d\Omega<0.26$ $\mu$b/sr & 90\% & \cite{Shirotori:2012ka}\\
\end{tabular}
\end{ruledtabular}
\end{table}

\subsection{Photoproduction experiments}

A series of experiments were carried out at the Thomas Jefferson National Accelerator Facility with the CLAS detector and the Hall-B photon tagging system to search for the $\Theta^+$ baryon via exclusive photoproduction processes~\cite{Stepanyan:2003qr,McKinnon:2006zv,Niccolai:2006td,Kubarovsky:2003fi,Battaglieri:2005er,DeVita:2006ha}. In these experiments, the LD$_2$ target~\cite{Stepanyan:2003qr,McKinnon:2006zv,Niccolai:2006td} and the LH$_2$ target~\cite{Kubarovsky:2003fi,Battaglieri:2005er,DeVita:2006ha} were used, and the energy of the photon beam was of several GeV.

In the early $\gamma d$ experiment~\cite{Stepanyan:2003qr}, a peak was observed in the invariant mass spectrum $M_{nK^+}$ via the process $\gamma d\rightarrow pK^-K^+n$. The background from the $\phi$ meson and $\Lambda(1520)$ was suppressed by removing the events with $M_{K^+K^-}<1.07$ GeV/$c^2$ or $1.458<M_{pK^-}<1.551$ GeV/$c^2$. Besides, the neutron momentum $p_n$ was required to be larger than 80 MeV/$c$ and the $K^+$ momentum was required to be less than 1.0 GeV/$c$. Then the fitted mass and width of the resonance are $M=1542\pm5$ MeV/$c^2$ and $\Gamma=21$ MeV/$c^2$ with 43 counts. The statistical significance is $5.2\pm0.6\sigma$. However, this result was not reproduced in a high statistics experiment~\cite{McKinnon:2006zv} with the data collected in two months in early 2004. With similar cuts that the events with $M_{K^+K^-}<1.06$ GeV/$c^2$ or $1.495<M_{pK^-}<1.545$ GeV/$c^2$ were removed, no peak was observed in $M_{nK^+}$ spectrum. The upper limits on the total cross section of $\Theta^+$ production in $\gamma d\rightarrow pK^-\Theta^+$ and $\gamma n\rightarrow K^-\Theta^+$ are respectively estimated to be 0.3 nb and 3 nb at 1540 MeV/$c^2$ with 95\% C.L.. The process $\gamma d \rightarrow \Lambda\Theta^+\rightarrow\Lambda nK^+$ was also analyzed~\cite{Niccolai:2006td}. The upper limit on the total cross section is estimated to be 5 nb at 1540 MeV/$c^2$ with 95\% C.L..

In the early $\gamma p$ experiment~\cite{Kubarovsky:2003fi}, a peak was observed in the invariant mass spectrum $M_{nK^+}$ via the process $\gamma p\rightarrow \pi^+ K^-K^+n$. However, in the high statistics experiments~\cite{Battaglieri:2005er,DeVita:2006ha}, this resonance was reproduced neither in $\gamma p\rightarrow \overline{K}^0K^+n$ nor in $\gamma p\rightarrow\overline{K}^0K^0p$. The contaminations from the $\phi$ and $\Lambda$ as well as other resonances were removed in a similar way. No peak was observed in $M_{nK^+}$ or $M_{pK_S^0}$ spectrum, even though the cuts on specific angular ranges were used. The upper limits on the total cross section for the $\Theta^+$ production in these two processes are respectively estimated to be 0.8 nb and 1.5 nb at 1540 MeV/$c$ with 95\% C.L.. The ratio for the yield of the $\Theta^+$ to the $\Lambda(1520)$ is estimated to be less than 0.22\% with 95\% C.L.. This null result also contradicts with the result by the SAPHIR Collaboration~\cite{Barth:2003es} via the same process and in nearly the same kinematic regions.

Based on the same data collected by the CLAS detector~\cite{DeVita:2006ha}, a narrow resonance was observed via the interference with the $\phi$ meson~\cite{Amaryan:2011qc}. In the reaction $\gamma p\rightarrow pK_S^0K_L^0$, the $K_S^0$ was reconstructed from two pions and the $K_L^0$ was reconstructed from the missing mass of the $pK_S^0$ system. The events under the $\phi$ peak $M_{K_S^0K_L^0}=1.02\pm0.01$ GeV/$c^2$ were selected, and only the data with the photon energy below 2.6 GeV were used. Either by restricting the $M_{pK_S^0}$ region like less than 1.56 GeV or by selecting the $-(p_\gamma-p_{K_S^0})^2$ region like less than 0.45 GeV$^2$, a narrow resonance at 1543 MeV/$c^2$ with a significance of $5.3\sigma$ was observed in $K_S^0$-missing mass spectrum which corresponded to the $M_{pK_L^0}$. However this result was not approved by the CLAS Collaboration~\cite{Anghinolfi:2012yg}.

Apart from these experiments by the CLAS Collaboration, the FOCUS Collaboration also searched for the $\Theta^+$ in a photoproduction experiment based on the 1996-1997 fixed-target data at FermiLab~\cite{Link:2006yh}. The photon beam was obtain from 300 GeV electrons and positrons and impinged on the BeO targets. A total of 72 million $K_S^0$ events were selected. No evidence was found for the $\Theta^+$ in the $M_{pK_S^0}$ spectrum from 1470 to 2200 MeV/$c^2$. The upper limit at 95\% C.L. on the cross section ratio $\sigma(\Theta^+)\mathcal{B}(pK_S^0)/\sigma(K(892)^+)$ is 0.13\% for $\Gamma\approx0$ MeV and 0.33\% for $\Gamma=15$ MeV, and the upper limit on the cross section ratio $\sigma(\Theta^+)\mathcal{B}(pK_S^0)/\sigma(\Sigma(1385)^\pm)$ is 2.3\% for $\Gamma\approx0$ MeV and 5.7\% $\Gamma=15$ MeV. In the case that the momentum of the parent particles is above 25 GeV/$c$, the limits are stricter.

\subsection{Neutrino experiment}

After the ITEP's analysis on the old neutrino data for the $\Theta^+$~\cite{Asratyan:2003cb,Asratian:2005rt}, the NOMAD Collaboration performed another analysis on neutrino-nucleon ($\nu_\mu N$) interactions at CERN for the $\Theta^+$ via the $pK_S^0$ decay mode, but provided null result~\cite{Samoylov:2006sb}. The total sample was about 1.5 million with both charged and neutral current events.

In this analysis, the background was studies in three different ways, Monte Carlo events containing no $\Theta^+$, fake pairs by combining a $K_S^0$ and a proton from different events and a polynomial fit to the $M_{pK_S^0}$ distribution excluding the $\Theta^+$ mass region. No signal for the $\Theta^+$ was observed. The upper limit on the $\Theta^+$ production rate is estimated as $2.13\times10^{-3}$ events per neutrino interaction at 1530 MeV/$c^2$ with 90\% C.L.. As claimed in Ref.\cite{Samoylov:2006sb}, two events used in the fake pair technique should have similar hadronic jet momenta, and therefore the background was underestimated in the ITEP's analysis~\cite{Asratyan:2003cb,Asratian:2005rt}.

\subsection{$e^+e^-$ experiments}

The $\Theta^+$ baryon was searched for at $e^+e^-$ colliders via quarkonium resonance processes~\cite{Bai:2004gk,Aubert:2005qi,Wang:2005fc,Aubert:2007qea}, $Z$ resonance processes~\cite{Schael:2004nm,Abdallah:2007ad}, continuum $q\bar{q}$ processes~\cite{Aubert:2005qi} and two-photon collisions~\cite{Achard:2006be}. No evidence was observed in any of these experiments.

The BES Collaboration analyzed 14 million $\psi(2S)$ and 58 million $J/\psi$ events to search for the $\Theta^+$ through both the $pK_S^0$ and the $nK^+$ decay modes~\cite{Bai:2004gk}. By counting the number of events within the region $1520-1560$ MeV/$c^2$, the upper limits on the branching ratios of $\psi(2S)$ and $J/\psi$ decay modes including the $\Theta^+$ or the $\overline{\Theta}^-$ are set at 90\% C.L.:
\begin{equation}
\begin{split}\label{bes}
\mathcal{B}(\psi(2S)\rightarrow\Theta^+\overline{\Theta}^-\rightarrow K_S^0pK^-\bar{n}+K_S^0\bar{p}K^+n)&<0.84\times10^{-5},\\
\mathcal{B}(\psi(2S)\rightarrow\Theta^+K^-\bar{n}\rightarrow K_S^0pK^-\bar{n})&<1.0\times10^{-5},\\
\mathcal{B}(\psi(2S)\rightarrow\overline{\Theta}^-K^+n\rightarrow K_S^0\bar{p}K^+n)&<2.6\times10^{-5},\\
\mathcal{B}(\psi(2S)\rightarrow K_S^0p\overline{\Theta}^-\rightarrow K_S^0pK^-\bar{n})&<0.60\times10^{-5},\\
\mathcal{B}(\psi(2S)\rightarrow\Theta^+K_S^0\bar{p}\rightarrow K_S^0\bar{p}K^+n)&<0.70\times10^{-5},\\
\mathcal{B}(J/\psi\rightarrow\Theta^+\overline{\Theta}^-\rightarrow K_S^0pK^-\bar{n}+K_S^0\bar{p}K^+n)&<1.1\times10^{-5},\\
\mathcal{B}(J/\psi\rightarrow\Theta^+K^-\bar{n}\rightarrow K_S^0pK^-\bar{n})&<2.1\times10^{-5},\\
\mathcal{B}(J/\psi\rightarrow\overline{\Theta}^-K^+n\rightarrow K_S^0\bar{p}K^+n)&<5.6\times10^{-5},\\
\mathcal{B}(J/\psi\rightarrow K_S^0p\overline{\Theta}^-\rightarrow K_S^0pK^-\bar{n})&<1.1\times10^{-5},\\
\mathcal{B}(J/\psi\rightarrow\Theta^+K_S^0\bar{p}\rightarrow K_S^0\bar{p}K^+n)&<1.6\times10^{-5}.
\end{split}
\end{equation}

The BaBar Collaboration analyzed the data at the PEP-II asymmetric-energy $e^+e^-$ storage ring at SLAC with the center-of-mass energy at or just below the $\Upsilon(4S)$ resonance to search for the $\Theta^+$ via the $pK_S^0$ decay mode~\cite{Aubert:2005qi}. A total of 2.7 million $K_S^0$ events were selected. By fitting the $M_{pK_S^0}$ spectrum with fixed $\Theta^+$ mass values in the region 1520-1550 MeV/$c^2$, no hint for a signal was found. The upper limit on the total cross section is 171 fb for $\Gamma=1$ MeV and 363 fb for $\Gamma=8$ MeV. The upper limit on the yield per $q\bar{q}$ event is $5.0\times10^{-5}$ for $\Gamma=1$ MeV and $11\times10^{-5}$ for $\Gamma=8$ MeV, and the upper limit on the yield per $\Upsilon(4S)$ event is $18\times10^{-5}$ for $\Gamma=1$ MeV and $37\times10^{-5}$ for $\Gamma=8$ MeV. All these upper limits are set at 95\% C.L. with the assumption that $\mathcal{B}(\Theta^+\rightarrow pK_S^0)=25\%$.

The Belle Collaboration analyzed 152 million $B\overline{B}$ decays at the KEKB asymmetric-energy $e^+e^-$ collider at the $\Upsilon(4S)$ resonance~\cite{Wang:2005fc}. The $\Theta^+$ was searched for in the three-body decay $B^0\rightarrow p\bar{p}K_S^0$. No peak was observed in the $M_{pK_S^0}$ spectrum in the region 1530-1550 MeV/$c^2$. The upper limit on the branching ratio is $\mathcal{B}(B^0\rightarrow\Theta^+\bar{p}\rightarrow p\bar{p}K_S^0)<2.3\times10^{-7}$ at 90\% C.L.. This process was also studied by the BaBar Collaboration with 232 million $B\overline{B}$ decays~\cite{Aubert:2007qea}. No resonance was observed in the mass region 1520-1550 MeV/$c^2$ either. The upper limit on the branching ratio is improved to $0.5\times10^{-7}$ at 90\% C.L..

The ALEPH Collaboration searched for the $\Theta^+$ in four million hadronic $Z$ decays recorded by the ALEPH detector during the LEP1 operation in the years 1991-1995~\cite{Schael:2004nm}. A total of 480 000 events were selected to search for a narrow resonance in the $M_{pK_S^0}$ spectrum from 1500 to 1600 MeV/$c^2$, but no signal was observed. The upper limit on the $\Theta^+$ yield per hadronic $Z$ decay is $2.5\times10^{-3}$ at 95\% C.L. with the assumption that $\mathcal{B}(\Theta^+\rightarrow pK_S^0)=25\%$. A similar analysis was performed by the DELPHI Collaboration with the data recorded by the DELPHI detector during the LEP1 operation in the years 1991-1995~\cite{Abdallah:2007ad}. The upper limit on the $\Theta^+$ yield per hadronic $Z$ decay is $2.0\times10^{-3}$ at 95\% C.L. with the same assumption for the branching ratio to $pK_S^0$.

The L3 Collaboration searched for the $\Theta^+$ through the two-photon collisions $e^+e^-\rightarrow e^+e^-\gamma\gamma\rightarrow e^+e^-+hadrons$~\cite{Achard:2006be}. About 1.3 million hadronic events were selected with the two-photon center-of-mass energy greater than 5 GeV, and about 140 000 $K_S^0$ events were selected to search for the $\Theta^+$. No evidence was observed in the $pK_S^0$ and $\bar{p}K_S^0$ mass spectrum. The upper limit on the yield of the $\Theta^+$ per hadronic two-photon event is $4.7\times10^{-3}$ at 95\% C.L. with the assumption that $\mathcal{B}(\Theta^+\rightarrow pK_S^0)=25\%$.

\subsection{$ep$ DIS experiment}

Another search for the $\Theta^+$ in the DIS experiment at HERA was performed by the H1 Collaboration~\cite{Aktas:2006ic} apart from the ZEUS experiment~\cite{Chekanov:2004kn}, but provided null result. This analysis was based on the data of 27.6 GeV electrons collided with 820 GeV protons in 1996-1997 and with 920 GeV protons in 1998-2000. The inelasticity $y$ was in the range 0.1-0.6. The $Q^2$ was divided into three intervals 5-10 GeV$^2$, 10-20 GeV$^2$ and 20-100 GeV$^2$. The $\Theta^+$ was search for in the $M_{pK_S^0}$ and $M_{\bar{p}K_S^0}$ spectrum from 1.48 to 1.7 GeV/$c^2$ with the width varying from 4.8 to 11.3 MeV/$^2$, but a narrow resonance was observed in none of the $Q^2$ bins. Even using the same cuts in~\cite{Chekanov:2004kn}, no significant signal appeared. The upper limit at 95\% C.L. on the production cross section is 120-360 pb depending on the mass between 1.48 and 1.7 GeV/$c^2$ with the assumption that $\mathcal{B}(\Theta^+\rightarrow pK_S^0)=25\%$.

\subsection{Hadron-induced reactions}

The $\Theta^+$ baryon was widely searched for in proton-proton~\cite{AbdelBary:2004ts,AbdelBary:2006wd,Nekipelov:2006if,Alt:2003vb} or proton-antiproton collisions~\cite{Litvintsev:2004yw}, proton- or deuteron-nucleus collisions~\cite{Abt:2004tz,Antipov:2004jz,Pinkenburg:2004ux,Longo:2004gd} and meson-nucleon collisions~\cite{Adamovich:2005ns,Napolitano:2004mn,Miwa:2007xk,Shirotori:2012ka,Abe:2005gy}, but no evidence was found in any of these experiments.

An exclusive measurement in $pp$ collisions for the $\Theta^+$ was first performed by the COSY-TOF Collaboration~\cite{AbdelBary:2004ts}. The data were taken from the production runs in 2000 and 2002 with a beam momentum of 2.95 GeV/$c$ injected into the LH$_2$ target. The $\Theta^+$ was searched for in the process $pp\rightarrow \Sigma^+K^0p$, where the $K^0$ was reconstructed from a $\pi^+\pi^-$ pair as the $K_S^0$ and the $\Sigma^+$ was reconstructed from the delayed decay into a $n\pi^+$ pair. A smooth background of the $M_{pK_S^0}$ distribution was obtained by a polynomial fit excluding the region from 1.51 to 1.54 GeV/$c^2$. A peak was observed at 1530$\pm$5 MeV/$c^2$ with a total cross section of $0.4\pm0.1\pm0.1$ $\mu$b.

However, the improved experiment by COSY-TOF with four or eight times higher statistics and a slightly higher beam momentum of 3.059 GeV/$c$ did not confirm this resonance~\cite{AbdelBary:2006wd}. The $\Sigma^+$ was reconstructed from the missing mass of the $pK_S^0$ system instead. The data was analyzed with three independent programs, but no peak was observed in the region from 1.50 to 1.55 GeV/$c^2$. The upper limit on the total cross section for the $\Theta^+$ production is derived to be 0.15 $\mu$b with 95\% C.L..

The $\Theta^+$ baryon was searched for in the reaction $pp\rightarrow pK^0\pi^+\Lambda$ at the research centre J\"{u}lich at COSY~\cite{Nekipelov:2006if}. The momentum of the proton beam was 3.65 GeV/$c$. In the analysis, the $\Lambda$ was reconstructed from the $\pi^-p$ decay mode, and then the $K^0$ meson was reconstructed from the missing mass of the $p\pi^+\pi^-p$ system. In the missing mass distribution of the $\pi^+\Lambda$ system, no peak around 1.54 GeV/$c^2$ was observed. With the mass varying from 1.52 to 1.54 GeV/$c^2$ and the width varying from 1 to 15 MeV/$c^2$, the upper limit on the total cross section of $pp\rightarrow\Theta^+\pi^+\Lambda$ is 58 nb at 95\% C.L.. This has no contradiction with the old result by the COSY-TOF Collaboration~\cite{AbdelBary:2004ts}, although the comparison may be questionable. In another fixed-target $pp$ collision with 158 GeV beam performed by the NA49 Collaboration at CERN, the resonance for the $\Theta^+$ in the $pK_S^0$ decay mode was not observed either~\cite{Alt:2003vb}.

By analyzing $p\bar{p}$ collision data recorded by the CDF \uppercase\expandafter{\romannumeral2} detector at Tevatron, the CDF Collaboration reported the result of searching for the $\Theta^+$ in the $pK_S^0$ mode~\cite{Litvintsev:2004yw}. Based on the data from different trigger paths, the upper limits on the $\Theta^+$ yield are 89 and 76 at 90\% C.L., while 172 and 312 events were required for the $3\sigma$ observation respectively.

Taking advantage of the a huge sample of 200 million minimum bias events of a fixed-target experiment at the 920 GeV proton storage ring of DESY, the HERA-B Collaboration searched for the $\Theta^+$ in the $pK_S^0$ decays~\cite{Abt:2004tz}. The data were taken at mid-rapidity on the carbon, titanium and tungsten targets during the 2002-2003 run. No evidence for a narrow resonance was observed in the mass region 1450-1700 MeV/$c^2$. The upper limit on the ratio for the yield of $\Theta^+$ to $\Lambda(1520)$ is 2.7\% at 95\% C.L..

Then based on more than 600 million data collected in March and April in 1999 with 70 GeV ITEP proton beam injected into the carbon target, the SPHINX Collaboration searched for the $\Theta^+$ via the $nK^+$, $pK_S^0$ and $pK_L^0$ decay modes in the reaction $pN\rightarrow\Theta^+\overline{K}^0N$~\cite{Antipov:2004jz}. The $\Theta^+\overline{K}^0$ was detected in four final states $(nK^+)K_S^0$, $(pK_S^0)K_L^0$, $(pK_L^0)K_S^0$ and $(pK_S^0)K_S^0$. After the $\Lambda(1520)$ and $\phi$ cuts, no signal for the $\Theta^+$ was visible. The upper limits on the cross sections are 26, 42, 39 and 52 nb per nucleon respectively at 90\% C.L..

The PHENIX Collaboration searched for the anti-baryon $\overline{\Theta}^-$ via the $K^-\bar{n}$ channel by analyzing the $d$Au data at the nucleon center-of-mass energy of 200 GeV at RHIC~\cite{Pinkenburg:2004ux}. The anti-neutrons were detected through the annihilation signals. The momentum of the $K^-$ was required to be less than 1.5 GeV/$c$. After the timing correction applied, no statistically significant peak was observed around 1540 MeV/$c^2$.

The HyperCP Collaboration searched for the $\Theta^+$ in an experiment that a charged secondary beam produced by the 800 GeV protons interacting in a copper target~\cite{Longo:2004gd}. The beam with the momentum of 120-220 GeV/$c$ was a mixture of mainly protons and $\pi^+$ as well as about 5\% $K^+$. The events with $\Lambda\rightarrow p\pi^-$ or $K^+\rightarrow\pi^+\pi^+\pi^-$ were removed. In the selected sample of 106 000 $pK_S^0$ and $\bar{p}K_S^0$ events, no resonance was found around 1540 MeV/$c^2$. The upper limit on the ratio of the $\Theta^+$ events in the sample is 0.3\% at 90\% C.L..

Based on the data from the Large Acceptance Superconducting Solenoid (LASS) spectrometer facility at SLAC, the $\Theta^+$ was searched for in the reaction $K^+p\rightarrow K^+n\pi^+$ with 11 GeV/$c$ $K^+$ beam and the LH$_2$ target~\cite{Napolitano:2004mn}. With the assumption that $\mathcal{B}(\Theta^+\rightarrow nK^+)=50\%$, a narrow resonance for $\Theta^+$ was forbidden in this analysis.

Using the data of low momentum kaons secondary interacting with the material of the detector, the Belle Collaboration searched for the $\Theta^+$ in an inclusive and an exclusive processes~\cite{Abe:2005gy}. The data were collected at and 60 MeV below the $\Upsilon(4S)$ resonance at the KEKB. The probable momentum of the kaon was 0.6 GeV/$c$. For the inclusive production, no peak was observed in the $M_{pK_S^0}$ spectrum. The upper limit on the ratio of the yield of $\Theta^+$ to $\Lambda(1520)$ is 2.5\% at 90\% C.L. with the assumption that $\mathcal{B}(\Theta^+\rightarrow pK_S^0)=25\%$. For the exclusive reaction $K^+n\rightarrow pK_S^0$, no signal for the $\Theta^+$ was found either. The upper limit on the width $\Gamma(K^+n\rightarrow\Theta^+\rightarrow pK_S^0)$ estimated as Eq.(\ref{iwidth}) is 0.64 MeV/$c^2$ at $M=1539$ MeV/$c^2$ at 90\% C.L.. This upper limit does not contradict with the new result by the DIANA~\cite{Barmin:2009cz,Barmin:2013lva}.

The WA89 Collaboration performed the only $\Sigma^-$ hyperon induced experiment to search for the $\Theta^+$. The data were collected in 1993 and 1994 at CERN~\cite{Adamovich:2005ns}. The hyperon beam had a mean momentum of 340 GeV/$c$ and a momentum spread of $\sigma(p)/p=9\%$. The carbon and copper targets were used. Within the $pK_S^0$ sample of 5.2 million events, no resonance was found around the mass 1540 MeV/$c^2$. The upper limit on the production cross section times the branching ratio to $pK_S^0$ is 1.8 $\mu$b per nucleon in the region $x_F>0.05$ at 99\% C.L., and the limit on the differential cross section $d\sigma\cdot\mathcal{B}(pK_S^0)/dx_F$ is 18 $\mu$b for $0.05<x_F<0.15$.

The $\Theta^+$ was searched for via the reaction $K^+p\rightarrow\pi^+X$ in the E559 experiment at the K6 beam line of the KEK 12 GeV proton synchrotron~\cite{Miwa:2007xk}. The $K^+$ beam had a central momentum of 1.2 GeV/$c$. In total, the number of the selected irradiated $K^+$ was 3.31 billion collected in one month in June 2005 and 2.17 billion collected in two weeks in December 2005. The LH$_2$ target was used. In the missing mass spectrum of the $\pi^+$, no peak was observed. The upper limit on the differential cross section averaged over 2-22$\degree$ in the laboratory frame is 3.5 $\mu$b/sr at 90\% C.L..

The $\Theta^+$ was searched for in a very recent E19 experiment with the K1.8 beam line in the Hadron Facility at the Japan Proton Accelerator Research Complex (J-PARC) via the reaction $\pi^-p\rightarrow K^-X$~\cite{Shirotori:2012ka}.  The $\pi^-$ beam with the momentum of 1.92 GeV/$c$ was incident on the LH$_2$ target. No resonance for the$\Theta^+$ was observed in the missing mass spectrum in the region 1510-1550 GeV/$c^2$.  The upper limit on the differential cross section averaged over 2-15$\degree$ in the laboratory frame is 0.26 $\mu$b/sr at 90\% C.L..

\section{The non-strange pentaquark state}

If the $S=+1$ pentaquark state $\Theta^+$ exists, the existence of a $SU(3)$ flavor multiplet is implied. Since the mass of the observed $\Theta^+$ candidate is close to the predicted value of the lightest member of the antidecuplet in the chiral soliton model~\cite{Diakonov:1997mm}, the non-strange members in the antidecuplet are of high interest. The neutral state $N^*$ was predicted to be a $P_{11}$ nucleon-like resonance around 1680 MeV/$c^2$ with a narrow width. As suggested in the Ref.~\cite{Polyakov:2003dx}, it can be probed in the reactions $\gamma n\rightarrow\eta n$ and $\gamma n\rightarrow K\Lambda$. Besides an enhancement of the cross section of the reaction $\gamma n\rightarrow\eta n$ with respect to that of the reaction $\gamma p\rightarrow\eta p$ was predicted in this model.

With the analysis of the data collected at the GRAAL facility in 2002, a candidate for the $N^*$ was reported~\cite{Kuznetsov:2004gy,Kuznetsov:2007gr}. In this analysis, both the quasi-free $\gamma n\rightarrow\eta n$ and $\gamma p\rightarrow\eta p$ were explored using a deuteron target. The $\eta$ was reconstructed via two photons. The recoil nucleons emitted at forward angles with $\theta_\textrm{lab}<23\degree$ and emitted at central angles with $\theta_\textrm{lab}>26\degree$ were detected. A cut on the missing mass of the $\eta$ was imposed to reject the background from the reaction $\gamma N\rightarrow\eta XN$. For the $\gamma n$ reaction, a bump was observed in both the invariant mass $M_{\eta n}$ and the photon nucleon center-of-mass energy $W$ spectra near 1.68 GeV/$c^2$. Its width is estimated to be less than 30 MeV/$c^2$. In contrast, the bump does not appear in the data of the $\gamma p$ reaction.

Then this resonance was also measured in the experiment performed at the LNS at Tohoku University~\cite{Miyahara:2007zz}. In this experiment, the photon beam with the energy from 0.6 to 1.15 GeV were injected into the deuteron target. For the reaction $\gamma N\rightarrow\eta N$, the events with three neutral particles were selected. The $\eta\pi N$ events were removed by momentum difference of the $\eta$ in the center-of-mass system. The cross section for the process $\gamma d\rightarrow\eta pn$ had an enhancement at $E_\gamma\sim1010$ MeV. By subtracting the contribution from $\gamma p\rightarrow\eta p$ estimated from the $\eta$ production cross section on the proton including its momentum distribution in the deuteron, the enhancement corresponds to a bump in the cross section for $\gamma n\rightarrow \eta n$. The mass and the width are deduced as $M=1666\pm5$ MeV/$c^2$ and $\Gamma\leq40$ MeV/$c^2$.

The CBELSA/TAPS Collaboration investigated the same reaction and supported the bump structure in the excitation function for the $\eta$ production off the neutron~\cite{Jaegle:2008ux}. In this analysis, the $\eta$ was reconstructed via the decay $\eta\rightarrow3\pi^0\rightarrow6\gamma$. Then the events with at least six neutral hits which could be combined to three pions were selected. The recoil nucleons were treated as missing particles, and the missing masses were calculated under the assumption of quasi-free meson production on a nucleon at rest. In the data for the neutron, a bump in the cross section was observed around photon energy of 1 GeV. The fitted mass and width of this structure with the Breit-Wigner form are $M=1683$ MeV/$c^2$ and $\Gamma=60\pm20$ MeV/$c^2$. In a new analysis, the effects of the Fermi motion were removed~\cite{Jaegle:2011sw}. The fitted mass and width are $M=1663\pm3$ MeV/$c^2$ and $\Gamma=25\pm12$ MeV/$c^2$.

This structure was studied at the Mainz MAMI accelerator facility either~\cite{Werthmuller:2010af}. The events of the quasi-free reaction $\gamma p\rightarrow\eta p$ were identified with exactly two neutral hits coming from the $\eta$ decay and exactly one charged hit as the proton. For the quasi-free reaction on the neutron, the events were selected with exactly three neutral hits with two of them combined to form an $\eta$ meson, and then the coplanar condition for the $\eta$ meson and the recoil nucleon was imposed by a cut on their relative azimuthal angle as $130\degree<|\Delta\phi|<220\degree$ or more strict as $170\degree<|\Delta\phi|<190\degree$. The background from the $\eta\pi N$ final states was remove by a cut on the missing mass of the $\eta$ meson calculated under the assumption of a free nucleon at rest. Then a peak appeared in the excitation function with respect to the $M_{\eta n}$, while no corresponding peak appeared in the quasi-free production on the proton case. The fitted mass and width with the Breit-Wigner form are $M=1675$ MeV/$c^2$ and $\Gamma=30\pm10$ MeV/$c^2$.

Besides, the $N^*$ was also searched for in the quasi-free Compton scattering on the neutron~\cite{Kuznetsov:2010as}. The data were collected at the GRAAL facility. In this analysis, the reactions $\gamma n\rightarrow\gamma n$, $\gamma n\rightarrow\pi^0n$, $\gamma p\rightarrow\gamma p$ and $\gamma p\rightarrow\pi^0p$ were investigated. The coplanar condition and the missing mass cut were applied to identify the $\gamma N$ final states but still contaminated by the $\pi^0$ production events. Using the cut on the missing energy $E_\textrm{miss}=E_\gamma-E_{\gamma'}-T_N$, where $T_N$ is the kinetic energy of the recoil nucleon, the Compton scattering events and the $\pi^0$ production events were further selected with $-0.05\leq E_\textrm{miss}\leq0.05$ GeV and $0.07\leq E_\textrm{miss}\leq0.15$ GeV respectively. In the distribution of the Compton scattering events on the neutron with respect to the center-of-mass energy $W$, a peak was observed at $M=1686\pm7\pm5$ MeV/$c^2$ with a narrow width $\Gamma<30$ MeV/$c^2$.

In all the above experiments, the $N^*$ pentaquark candidate was produced via the photoexcitation on the neutron. Whereas, accounting for the $SU(3)$ flavor symmetry violation, the photoexcitation can also occur on the proton, even if it may be highly suppressed. With the data collected at the GRAAL facility in 1998-1999, a narrow resonance in the region from 1550 to 1750 MeV/$c^2$ was searched for through the analysis on the beam asymmetry for the $\eta$ meson photoproduction on the free protons~\cite{Kuznetsov:2007dy,Kuznetsov:2008hj,Kuznetsov:2008ii}. The photon polarization degree varied from 0.5 to 0.85 depending on the energy. The events with two photons detected in the BGO Ball were initially selected by the invariant mass of the $\eta$ reconstructed from the two photon, and the events with one photon detected in the forward wall and the other photon detected in the BGO Ball were initially selected using the missing mass calculated from the measured recoil proton. In the asymmetry distribution, a peak appeared at the forward angles and was replaced by an oscillating structure at central angles. With the narrow resonances $S_{11}$, $P_{11}$, $P_{13}$ and $D_{13}$ added one by one with the Breit-Wigner form, either the $P_{11}$, $P_{13}$ or $D_{13}$ improved the description of the data. The mass and width of the resonance is $M=1688\pm2\pm5$ MeV/$c^2$ and $\Gamma\leq15$ MeV/$c^2$.

\section{Other pentaquark candidates}

\subsection{The $\Theta^{++}$}

The $\Theta^{++}$ baryon, if exists, is recognized as a possible isospin partner of the $\Theta^+$, if the $\Theta^+$ is not a isospin singlet. It was predicted in some chiral soliton models~\cite{Wu:2003mc,Wu:2003xc,Wu:2004ca}. The lowest constituent state of the $\Theta^{++}$ is $|uuud\bar{s}\rangle$. Thus the $pK^+$ is estimated as the dominant decay channel.

The only signal for the $\Theta^{++}$ was reported by the STAR Collaboration at RHIC~\cite{Huang:2005nk}. By analyzing 18.6 million minimum bias $d$Au collisions at the nucleon center-of-mass energy of 200 GeV, a narrow peak was observed at $1528\pm2\pm5$ GeV/$c^2$ in the $pK^++\bar{p}K^-$ spectrum. The background was estimated by one kaon and one proton from two events with similar multiplicity and nearby primary vertex location. Then the statistical significance is about $4.2\sigma$. The yield ratio to the $\Lambda(1520)$ is estimated to be 0.4\%. In the AuAu and CuCu collision data, nevertheless, this peak is neither confirmed nor ruled out. However, this resonance was not observed in any other experiment as listed in the Table \ref{theta++}.
\begin{table}
\caption{Experimental search for the $\Theta^{++}$.}
\begin{ruledtabular}
\begin{tabular}{lccccr}\label{theta++}
Group & Reaction & Result & Confidence & Reference\\
\hline
STAR & $d\textrm{Au}\rightarrow pK^+(\bar{p}K^-)X$ & $M=1528$ MeV & $4.2\sigma$ & \cite{Huang:2005nk}\\
\hline
CLAS & $\gamma p\rightarrow pK^+K^-$ & No evidence & -- & \cite{Kubarovsky:2003fi,Juengst:2003yy}\\
CLAS & $\gamma p\rightarrow pK^+K^-$ & $\sigma<0.15$ nb, $\Gamma<0.1$ MeV & 95\% & \cite{Kubarovsky:2006ns}\\
HERMES & $\gamma d\rightarrow pK^+X$ & Zero counts & 91\% & \cite{Airapetian:2003ri}\\
ZEUS & $e^\pm p\rightarrow e^\pm pK^+(\bar{p}K^-)X$ & No evidence & -- & \cite{Chekanov:2004hd}\\
DELPHI & $Z\rightarrow pK^+X$ & $\langle N(\Theta^{++})\rangle<1.6\times10^{-3}$ & 95\% & \cite{Abdallah:2007ad}\\
BaBar & $B^+\rightarrow p\bar{p}K^+$ & $\mathcal{B}(\Theta^{++})\cdot\mathcal{B}(pK^+)\lesssim10^{-7}$ & 90\% & \cite{Aubert:2005gw}\\
HALL-A & $ep\rightarrow eK^-X$ & $N(\Theta^{++})/N(\Lambda(1520))<1.1\%$ & 90\% & \cite{Qiang:2006tp}\\
\end{tabular}
\end{ruledtabular}
\end{table}

\begin{table}
\caption{Experimental search for the $\Theta_c^0$.}
\begin{ruledtabular}
\begin{tabular}{lccccr}\label{thetac}
Group & Reaction & Result & Confidence & Reference\\
\hline
H1 & $ep\rightarrow eD^{*-}pX$ & $M=3099$ MeV & $5.4\sigma$ & \cite{Aktas:2004qf} \\
\hline
ZEUS & $e^\pm p\rightarrow e^\pm D^{*+}\bar{p}(D^{*-}p)X$ & $N(\Theta_c^0)/N(D^*)<0.23\%$ & 95\% & \cite{Chekanov:2004qm}\\
CDF & $p\bar{p}\rightarrow D^{*-}pX$ & $N(\Theta_c^0)<21$ & 90\% & \cite{Litvintsev:2004yw}\\
ALEPH & $Z\rightarrow D^{*-}pX$ & $\langle N(\Theta_c^0)\rangle\cdot\mathcal{B}(D^{*-}p)<6.3\times10^{-4}$ & 95\% &    \\
    & $Z\rightarrow D^{-}pX$ & $\langle N(\Theta_c^0)\rangle\cdot\mathcal{B}(D^{-}p)<31\times10^{-4}$ & 95\% & \cite{Schael:2004nm}\\
DELPHI & $Z\rightarrow D^{*+}\bar{p}X$ & $\langle N(\Theta_c^0)\rangle\cdot\mathcal{B}(D^{*+}\bar{p})<8.8\times10^{-4}$ & 95\% & \cite{Abdallah:2007ad}\\
BaBar & $e^+e^-\rightarrow q\bar{q}\rightarrow D^{*-}pX$ & $\langle N(\Theta_c^0)\rangle\cdot\mathcal{B}(D^{*-}p)<3.4\times10^{-5}$ & 95\% & \cite{Aubert:2006qu}\\
    & $B^0\rightarrow D^{*-}p\bar{p}\pi^+$ & $\mathcal{B}(\Theta_c^0)\cdot\mathcal{B}(D^{*-}p)<14\times10^{-6}$ & 90\% &     \\
    & $B^0\rightarrow D^-p\bar{p}\pi^+$ & $\mathcal{B}(\Theta_c^0)\cdot\mathcal{B}(D^{-}p)<9\times10^{-6}$ & 90\% & \cite{Aubert:2006qx}\\
FOCUS & $\gamma\textrm{BeO}\rightarrow D^{*-}pX$ & $\langle N(\Theta_c^0)\rangle\cdot\mathcal{B}(D^{*-}p)<4.2\times10^{-4}$ & 95\% &    \\
    & $\gamma\textrm{BeO}\rightarrow D^{-}pX$ & $\langle N(\Theta_c^0)\rangle\cdot\mathcal{B}(D^{-}p)<5.0\times10^{-4}$ & 95\% & \cite{Link:2005ti}\\
\end{tabular}
\end{ruledtabular}
\end{table}

\begin{table}
\caption{Experimental search for the $\Xi(1860)$.}
\begin{ruledtabular}
\begin{tabular}{lccccr}\label{xi1860}
Group & Reaction & Result & Confidence & Reference\\
\hline
NA49 & $pp\rightarrow\Xi^-\pi^\pm X$ & $M_{\Xi^{--}}=1862$ MeV & $5.8\sigma$ & \cite{Alt:2003vb}\\
    &   & $M_{\Xi^0}=1864$ MeV &    &   \\
\hline
CLAS & $\gamma p\rightarrow K^+K^+\pi^+X$ & No evidence & -- & \cite{Price:2004hr}\\
CLAS & $\gamma d\rightarrow \Xi^-\pi^-X$ & $\sigma<0.7$ nb & 90\% & \cite{CLAS:2011aa}\\
HERMES & $\gamma d\rightarrow\Xi^-\pi^-X$ & $\sigma<2.1$ nb & 90\% & \cite{Airapetian:2004mi}\\
H1 & $ep\rightarrow e\Xi^-\pi^\pm X$ & $N(\Xi^{--})/N(\Xi(1530)<15-45\%$ & 95\% &   \\
    &   & $N(\Xi^0)/N(\Xi(1530)<10-50\%$ & 95\% & \cite{Aktas:2007dd}\\
ZEUS & $e^\pm p\rightarrow e^\pm \Xi\pi X$ & $N(\Xi(1860))/N(\Xi(1530))<29\%$ & 95\% & \cite{Chekanov:2005at}\\
HERA-B & $p\textrm{C}\rightarrow \Xi^-\pi^-X$ & $\mathcal{B}(\Xi^-\pi^-)\cdot N(\Xi^{--})/N(\Xi^-)<3\%$ & 95\% &    \\
    &   & $\mathcal{B}(\Xi^-\pi^-)\cdot N(\Xi(1530)<4\%$ & 95\% & \cite{Abt:2004tz}\\
WA89 & $\Sigma^-\textrm{C}(\textrm{Cu})\rightarrow \Xi^-\pi^-X$ & $\sigma<3.1(3.5)$ $\mu$b & 99\% & \cite{Adamovich:2004yk}\\
CDF & $p\bar{p}\rightarrow \Xi^-\pi^\pm X$ & $N(\Xi^{--})\cdot\mathcal{B}(\Xi^-\pi^-)/N(\Xi(1530))<1.7\%$ & 90\% & \\
    &   & $N(\Xi^0)\cdot\mathcal{B}(\Xi^-\pi^+)/N(\Xi(1530))<3.2\%$ & 90\% & \cite{Abulencia:2006zs}\\
STAR & $d\textrm{Au}\rightarrow \Xi^-\pi^-X$ & No evidence & -- & \cite{Huang:2005nk}\\
ALEPH & $Z\rightarrow \Xi^-\pi^\pm X$ & $\langle N(\Xi(1860)^{--})\rangle\cdot\mathcal{B}(\Xi^-\pi^-)<4.5\times10^{-4}$ & 95\% &     \\
    &   & $\langle N(\Xi(1860)^0)\rangle\cdot\mathcal{B}(\Xi^-\pi^+)<8.9\times10^{-4}$ & 95\% & \cite{Schael:2004nm}\\
DELPHI & $Z\rightarrow\Xi^-\pi^-X$ & $\langle N(\Xi(1860)^{--})\rangle\cdot\mathcal{B}(\Xi^-\pi^-)<2.9\times10^{-4}$ & 95\% & \cite{Abdallah:2007ad}\\
BaBar & $e^+e^-\rightarrow q\bar{q}\rightarrow \Xi^-\pi^-X$ & $\langle N(\Xi^{--})\rangle<0.74\times10^{-5}$ & 95\% & \cite{Aubert:2005qi}\\
FOCUS & $\gamma\textrm{BeO}\rightarrow\Xi^-\pi^-X$ & $N(\Xi^{--})\cdot\mathcal{B}(\Xi^-\pi^-)/N(\Xi(1530))<3.2\%$ & 95\% & \cite{Link:2007ab}\\
COMPASS & $\gamma\textrm{LiD}\rightarrow\Xi^-\pi^- X$ & $N(\Xi^{--})<79$ & 99\% & \cite{Ageev:2005ud}\\
E690 & $pp\rightarrow\xi^-\pi^\pm X$ & $N(\Xi^{--})/N(\Xi^-\pi^-)<0.3\%$ (1.3\% for $\overline{\Xi}^{++}$) & 95\% &   \\
    &   & $N(\Xi^0)/N(\Xi^-\pi^+)<1.1\%$ (1.3\% for $\overline{\Xi}^0$) & 95\% & \cite{Christian:2005kh}\\
\end{tabular}
\end{ruledtabular}
\end{table}

\subsection{The $\Theta_c^0$}

The $\Theta_c^0$ baryon, if exists, is the lightest charmed pentaquark state like the $\Theta^+$ but the $\bar{s}$ quark replaced with the $\bar{c}$ quark. Its lowest Fock state is $|uudd\bar{c}\rangle$ which has the same constituent quarks with the combination of $D^{*-}p$ or $D^-p$. Thus these are estimated as the dominant decay channels.

The signal for the $\Theta_c^0$ was only observed in the DIS experiment by the H1 Collaboration~\cite{Aktas:2004qf}. The analysis was based on the data at HERA in 1996-2000. The $D^*$ was reconstructed via the decay channal $D^*\rightarrow D^0\pi_s\rightarrow K\pi\pi_s$. In the distribution of $M_{D^*p}=M_{K\pi\pi_sp}-M_{K\pi\pi_s}+M_{D^*}$ with opposite-charge combinations, a peak was observed at $3099\pm3\pm5$ MeV/$c^2$ with a Gaussian width of $12\pm3$ MeV/$c^2$. The background was estimated from the Monte Carlo simulation. The number of the signal events is $50.6\pm11.2$, corresponding to a statistical significance of 5.4$\sigma$. However this resonance was not observed in any other experiment as listed in the Table \ref{thetac}.

\subsection{The $\Xi(1860)^{--}$}

The $\Xi(1860)^{--}$ baryon, if exists, is a double strangeness $S=-2$ pentaquark candidate, having the lowest Fock state $|ddss\bar{u}\rangle$. It is recognized as an isospin quartet together with its partners $\Xi_{3/2}^-$, $\Xi_{3/2}^0$ and $\Xi_{3/2}^+$. The primary decay channel of $\Xi^{--}$ is estimated to be $\Xi^-\pi^-$.

The signal for the $\Xi^{--}$ was only reported by the NA49 Collaboration at CERN~\cite{Alt:2003vb}. The analysis was base on the data of 158 GeV/$c$ proton beam colliding with the LH$_2$ target. The $\Xi^-$ was reconstructed via the decay channel $\Xi^-\rightarrow\Lambda\pi^-\rightarrow p\pi^-\pi^-$. Then 1640 $\Xi^-$ and 551 $\overline{\Xi}^+$ were selected. A peak was observed in the $\Xi^-\pi^-$ mass spectrum. Combined with the $\overline{\Xi}^+\pi^+$ data as the antiparticle, the fitted peak is at $1862\pm2$ MeV/$c^2$ with 69 signals over the background of 75 events corresponding to a statistical significance of 5.8$\sigma$ estimated as $S/\sqrt{S+B}$. One of its isospin partners $\Xi^0$ was also observed, and the fitted mass is $1864\pm5$ MeV/$c^2$. Unfortunately this resonance was not observed in any other experiment as listed in the Table \ref{xi1860}.

\section{Discussions}

\begin{figure}
\includegraphics[width=0.7\textwidth]{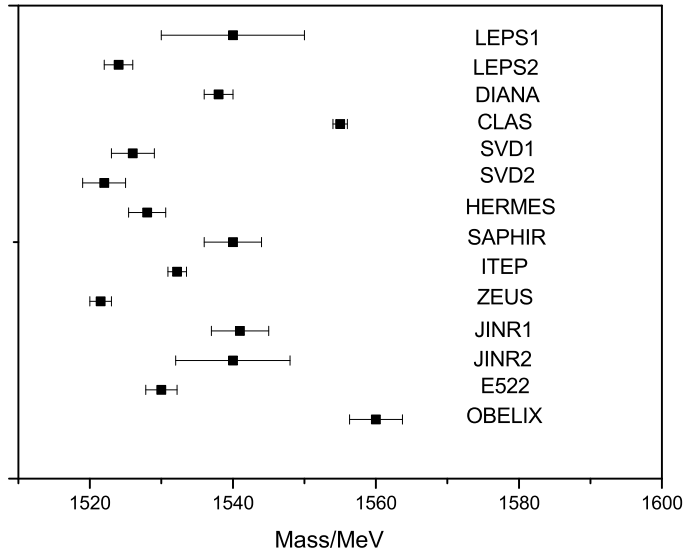}
\caption{\label{mass}Mass values for the $\Theta^+$ observed in various experiments. The error bars represent the statistical uncertainties.}
\end{figure}

\begin{figure}
\includegraphics[width=0.7\textwidth]{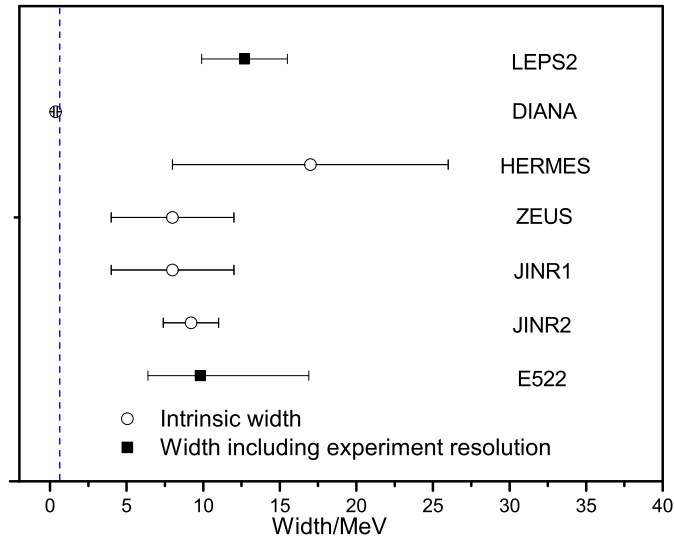}
\caption{\label{width}Width values for the $\Theta^+$ measured in various experiments. The error bars represent the statistical uncertainties. The experiments with only the upper limit on the width are listed in the Table \ref{thetay}. The dashed line is the upper limit of the intrinsic width of $\Theta^+$ at 90\% C.L. by Belle~\cite{Abe:2005gy}.}
\end{figure}

Among all the experiments in which the $\Theta^+$ was observed, the width was claimed to be narrow. However, the mass position of the signals in different experiments spreads in a large region from 1520 to 1560 MeV/$c^2$ as shown in Fig. \ref{mass}. This does not consist with a narrow resonance. The mass values in LEPS early experiment~\cite{Nakano:2003qx}, DIANA experiments~\cite{Barmin:2003vv,Barmin:2006we,Barmin:2009cz,Barmin:2013lva}, SAPHIR experiment~\cite{Barth:2003es} and JINR experiments~\cite{Troyan:2004wp,Aslanyan:2004gs} are around 1540 MeV/$c^2$, while the values in HERMES experiment~\cite{Airapetian:2003ri}, SVD early experiment~\cite{Aleev:2004sa}, ITEP's analysis on the old neutrino experiments~\cite{Asratyan:2003cb,Asratian:2005rt} and E522 experiment~\cite{Miwa:2006if} are near 1530 MeV/$c^2$. Besides, the LEPS recent experiment~\cite{Nakano:2008ee}, the SVD updated experiment~\cite{Aleev:2005yq} and the ZEUS experiment~\cite{Chekanov:2004kn} provide even lower mass values close to 1520 MeV/$c^2$, and the OBELIX experiment gives a much higher value. Therefore, it is possible that even the observed signals do not correspond to the same particle. The width of the $\Theta^+$ are not measured as accurate as the mass, and thus the values in almost all the experiments are consistent as shown in Fig. \ref{width}.

The statistical significance of the signals was usually estimated as $S/\sqrt{B}$ in the early experiments. This estimator neglects the uncertainty of the background, and thus the significance may be overestimated. The estimator $S/\sqrt{S+B}$ which assumes a smooth background with a well defined shape and $S/\sqrt{S+2B}$ which assumes a statistical independent background with uncertainties are more proper since the production mechanism is still unknown. If so, the significance of the signals in the SVD updated analysis~\cite{Aleev:2005yq} reduces to 5.6$\sigma$ and $4.1\sigma$ estimated as $S/\sqrt{S+2B}$ for the two samples respectively, but is still large enough as an evidence. In this case, however, the HERMES result decreases to 2.7-3.9$\sigma$, the JINR $p\textrm{C}_3\textrm{H}_8$ result decreases to 4.1$\sigma$ and the JINR $np$ result decreases to 3.5$\sigma$. These are not enough to be claimed as an evidence. Besides, the log-likelihood difference is also a suitable alternative and was used by the LEPS~\cite{Nakano:2008ee} and the DIANA~\cite{Barmin:2013lva}.

The $\Lambda$ produced in the inclusive experiments may affect the result of the $\Theta^+$ reconstructed via the $pK_S^0$ mode. As pointed out by M. Zavertyaev~\cite{Zavertyaev:2003wv}, the decay $\Lambda(1115)\rightarrow\pi^-p$ could lead to a spurious sharp peak at 1540 MeV/$c^2$ when the momentum of the $\Lambda$ is greater than 2 GeV/$c$. On the other hand, the $\pi^-p$ decays of the $\Lambda$s could enhance the background when the $\pi^-$ or the proton was paired with a $\pi^+$ or a $K_S^0$.

Up to now, no positive result is reported for the $\Theta^+$ production in the $e^+e^-$ experiments. A possible way to understand these null results with no contradict with the positive ones is to assume that the $\Theta^+$ production cross section is highly suppressed in $e^+e^-$ annihilations. Using the quark constituent counting rules, A.I. Titov et al. estimated the ratio of $\Theta^+$ to $\Lambda(1520)$ production in the fragmentation region and showed the ratio decreases very fast with energy~\cite{Titov:2004wt}. This ratio is often applied to estimate the yield of the $\Theta^+$, because $\Lambda(1520)$ is a narrow resonance with similar mass to the $\Theta^+$ and is easily reconstructed. The low value of this ratio implies a very different production mechanism for $\Theta^+$ if it really exists.

For the DIS experiments performed at HERA~\cite{Chekanov:2004kn,Aktas:2006ic}, the ZEUS and the H1 provided opposite conclusions. These two experiments were almost in the same conditions and with the data collected during the same period, but even using the same cuts the H1 could not produce the $\Theta^+$ signal observed by the ZEUS. It is very confused, and probably the signal observed by the ZEUS is a statistical fluctuation.

For the photoproduction experiments, there is a contradiction in the $\gamma p$ experiments between the upper limit on the cross section given by the CLAS~\cite{Battaglieri:2005er,DeVita:2006ha} and the result reported by the SAPHIR~\cite{Barth:2003es} and a contradiction in the $\gamma d$ experiments between the upper limit on the cross section given by the CLAS~\cite{McKinnon:2006zv} and the result reported by the LEPS~\cite{Nakano:2008ee}. As claimed by the LEPS, the contradiction between the LEPS and CLAS results is due to the different measurements. If the $\Theta^+$ is preferably produced at the forward angles, the CLAS would possibly not detect the $K^-$ meson associated with the $\Theta^+$, because the most forward angle for the $K^-$ detection is about $20\degree$ for the CLAS while most acceptance is of forward $20\degree$ for the LEPS~\cite{Nakano:2008ee}. It may be a similar case for the contradiction between the CLAS and the SAPHIR. If this is true, it will be a suggestion on the angular distribution for the $\Theta^+$ production. In the experiments at higher energy the $\Theta^+$ productions will be boosted to a much more forward direction and thus escape the coverage of the detectors in most high energy experiments. This may be a possible explanation for the null results in most high energy experiments.

In the improved analysis by the DIANA~\cite{Barmin:2006we,Barmin:2009cz,Barmin:2013lva}, a very narrow intrinsic width for the $\Theta^+$ was estimated as $0.36\pm0.11$ MeV/$c^2$. This result passed the upper limits given by the Belle~\cite{Abe:2005gy}, the E559~\cite{Miwa:2007xk} and the J-PARC~\cite{Shirotori:2012ka}. Therefore there is no contradiction between these experiments, although opposite conclusions were reported for the existence of the $\Theta^+$.

Among all the experimental search for the $\Theta^+$ pentaquark candidate, those with negative results have higher statistics and are consequently more reliable in usual, but it is hard to prove that all the observed peaks are fakes or fluctuations. Especially, the updated results by the LEPS~\cite{Nakano:2008ee}, the DIANA~\cite{Barmin:2006we,Barmin:2009cz,Barmin:2013lva} and the SVD~\cite{Aleev:2005yq} can be claimed as strong evidence for the $\Theta^+$. So the existence of the $\Theta^+$ is still debatable. For the other pentaquark candidates such as $\Theta^{++}$, $\Xi^{--}$ and $\Theta_c^0$, the signals are much less reliable since none of them is confirmed in any other experiment.

\section{Summary}

We reviewed the experimental search for the pentaquark states during the last decade. Both the most widely studied candidate $\Theta^+$ and the other candidates like $\Theta^{++}$, $\Xi^{--}$ and $\Theta_c^0$ as well as the non-strange pentaquark candidates are included. Since the first observation of the pentaquark-like baryon state $\Theta^+$, this field has aroused much interest, but even the existence of the pentaquark is debatable up to now.

If the pentaquark really exists, it will open a new world of the QCD and hadron physics. In particular, if the $\Theta^+$ exists, its production mechanism is almost unknown and needs to be investigate, whether it is a pentaquark or not. Besides it will imply the existence of a flavor multiplet. If the pentaquark does not exist, the peaks observed in the experiments with positive results need a reasonable explanation. In addition, the contradictions between the experiments should be examined in details. This will improve the analysis method and raise the reliability of the result in future experiments.

In order to confirm or rule out the existence of the pentaquark, particularly the $\Theta^+$, comparisons between experiments in similar conditions are required. Among the experiments, the updated results by the LEPS~\cite{Nakano:2008ee}, the DIANA~\cite{Barmin:2006we,Barmin:2009cz,Barmin:2013lva} and the SVD~\cite{Aleev:2005yq} provide the best positive evidence for the $\Theta^+$. Therefore more experiments at medium energy may lead to a clear conclusion.

\begin{acknowledgments}
This work is supported by National Natural Science Foundation of China (Grants No.~11035003 and No.~11120101004).
\end{acknowledgments}


%

\end{document}